\documentclass[twocolumn,astrosymb]{aastex631}  
\usepackage{mathtools}

\usepackage{tikz} 
\usepackage{enumitem}
\usepackage{amsmath,amsthm,amssymb,amsfonts,bm}
\usepackage{listings}
\usepackage{graphics}
\usepackage{url}
\usepackage{appendix}
\usepackage{kbordermatrix}
\usepackage[super]{nth}
\usepackage{fontawesome}
\usepackage{soul}
\usepackage{textcomp}

\newcommand{\smin}{$S_{\textrm{min}}$}
\newcommand{\nfc}{$\textrm{N}_{\textrm{FC}}$}

\newcommand{\Ttwofive}{$\langle \textrm{T}_{\textrm{sz}} \rangle_{2500}$}

\newcommand{\pcatde}{\texttt{PCAT-DE}}
\newcommand{\Tpwxtwofive}{$\langle \textrm{T}_{\textrm{x,pw}}\rangle_{2500}$}

\newcommand{\ASZpmw}{$\textrm{A}_{\textrm{PMW}}^{\textrm{SZ}}$}
\newcommand{\ASZplw}{$\textrm{A}_{\textrm{PLW}}^{\textrm{SZ}}$}
\newcommand{\ASZpmwhat}{$\hat{\textrm{A}}_{\textrm{PMW}}^{\textrm{SZ}}$}
\newcommand{\ASZplwhat}{$\hat{\textrm{A}}_{\textrm{PLW}}^{\textrm{SZ}}$}
\newcommand{\ASZb}{$\textrm{A}_{b}^{\textrm{SZ}}$}
\newcommand{\nsrc}{N$_{\textrm{src}}$}
\newcommand{\planck}{\emph{Planck}}

\newcommand{\Bzero}{$B_{0}^b$}
\newcommand{\mjybm}{mJy beam$^{-1}$}

\begin{document}

\title{\textbf{\texttt{PCAT-DE}: Reconstructing point-like and diffuse signals in astronomical images using spatial and spectral information}}
\shorttitle{PCAT-DE}
\correspondingauthor{Richard M. Feder}
\email{rfederst@caltech.edu}

\author[0000-0002-9330-8738]{Richard M. Feder}
\affiliation{California Institute of Technology Division of Physics, Math, and Astronomy. 1200 East California Boulevard, Pasadena, CA 91125}

\author[0000-0002-0941-0407]{Victoria Butler}
\affiliation{Rochester Institute of Technology, 1 Lomb Memorial Drive, Rochester, NY 14623, USA}

\author[0000-0002-6939-9211]{Tansu Daylan}
\affiliation{Department of Astrophysical Sciences, Princeton University, 4 Ivy Lane, Princeton, NJ 08544}
\affiliation{LSSTC Catalyst Fellow}

\author[0000-0001-8132-8056]{Stephen K. N. Portillo}
\affiliation{DIRAC Institute, Department of Astronomy, University of Washington, 3910 15th Ave NE, Seattle, WA 98195}
\affiliation{Department of Mathematical and Physical Sciences, Concordia University of Edmonton, 7128 Ada Boulevard, Edmonton, AB, T5B 4E4 Canada}

\author[0000-0002-8213-3784]{Jack Sayers}
\affiliation{California Institute of Technology Division of Physics, Math, and Astronomy. 1200 East California Boulevard, Pasadena, CA 91125}

\author[0000-0002-9813-0270]{Benjamin J. Vaughan}
\affiliation{Rochester Institute of Technology, 1 Lomb Memorial Drive, Rochester, NY 14623, USA}

\author[0000-0001-9765-0197]{Catalina V. Zamora}
\affiliation{DIRAC Institute, Department of Astronomy, University of Washington, 3910 15th Ave NE, Seattle, WA 98195}

\author[0000-0001-8253-1451]{Michael Zemcov}
\affiliation{Rochester Institute of Technology, 1 Lomb Memorial Drive, Rochester, NY 14623, USA}
\affiliation{Jet Propulsion Laboratory, 4800 Oak Grove Drive, Pasadena, CA 91109, USA}

\begin{abstract}
Observational data from astronomical imaging surveys contain information about a variety of source populations and environments, and its complexity will increase substantially as telescopes become more sensitive. Even for existing observations, measuring the correlations between point-like and diffuse emission can be crucial to correctly inferring the properties of any individual component. For this task information is typically lost, either because of conservative data cuts, aggressive filtering or incomplete treatment of contaminated data. We present the code \texttt{PCAT-DE}, an extension of probabilistic cataloging designed to simultaneously model point-like and diffuse signals. This work incorporates both explicit spatial templates and a set of non-parametric Fourier component templates into a forward model of astronomical images, reducing the number of processing steps applied to the observed data. Using synthetic \emph{Herschel}-SPIRE multiband observations, we demonstrate that point source and diffuse emission can be reliably separated and measured. We present two applications of this model. For the first, we perform point source detection/photometry in the presence of galactic cirrus and demonstrate that cosmic infrared background (CIB) galaxy counts can be recovered in cases of significant contamination. In the second we show that the spatially extended thermal Sunyaev-Zel'dovich (tSZ) effect signal can be reliably measured even when it is subdominant to the point-like emission from individual galaxies. 
\end{abstract}

\section{Introduction}
The signal of interest in astronomical images is often contaminated by one or more other signals. These additional components can bias estimates of the desired signal when left unmodeled, and lower the precision with which we can infer correlated spatio-spectral structure. Estimating the effect of such components is a challenge, and mitigation strategies are situation-dependent.

Oftentimes the goal is to measure the emission from spatially unresolved sources (hereafter referred to as point sources) in the presence of diffuse signals, for example radio sources in front of the cosmic microwave background or behind galactic synchrotron \citep[][]{Hale2019}, or stars embedded in regions of high nebulosity \citep[][]{deBruijne2015}. The effects of diffuse structured signals can sometimes be mitigated using the fact that many diffuse astrophysical signals have angular power spectra that decrease with wavenumber. This motivates spatial (or angular) high pass filtering, either in real or Fourier space.  However, filtering approaches necessarily attenuate and distort the signal of interest, and often add uncertainties to signal estimates that can be difficult to assess. In the other limit, there are cases where the signal of interest is some type of diffuse structured emission and point sources are the contaminants. A common approach is to mask out known or suspected point source contaminants \citep[\textit{e.g.}][]{Barreiro2009},
but such approaches can be problematic when the spatial density of point sources and/or beam size necessitates removing a significant fraction of the image \citep[\textit{e.g.}][]{Zemcov2014}. Crucially, such removal is always to a finite detection limit, and the remaining point sources contaminate the estimate for the diffuse emission. This effect can be characterized, again at the cost of larger uncertainties on the signal of interest \citep[\textit{e.g.,}][]{Traficante2011}.

Many methods for separating point-like and diffuse signals exist. A review of source detection strategies is presented in \citet{Masias2012}; work in this field since this review includes \citet{Masias2015, Zheng2015, PORTILLO_17, Ofek2018, Robotham2018, Lukic2019, Collin2021, VILiu21, Du2022}. When spatial and/or spectral source properties are well understood, matched filtering is an effective method of source extraction \citep{Ofek2018, Lang20}, though optimal results only hold under strict assumptions, e.g., sources are isolated in background dominated images, with perfect knowledge of the PSF, noise model, etc. Multi-scale methods decompose images into components with fluctuation power on different spatial scales, enabling more reliable source detection and deblending in the presence of noise and structured backgrounds \citep{getsources, molinari11}. These approaches can involve several transformations of the data, meaning the quality of extraction of one component (typically the point sources) is emphasized at the cost of poor fidelity on the other components, however work has been done to improve signal reconstructions through more informed transformations and data representations \citep{dabbech15, ellien21}. It has been shown that point source photometry in the presence of nebulosity can be improved considerably by learning a pixel-wise, non-stationary covariance matrix for the structured signal surrounding each source (see \citealt{Saydjari22}; application to DECaPS2 survey in \citealt{decaps2}). However the use of a fixed input catalog in the post-processing step means that errors related to biases in source detection in the presence of diffuse signals are left uncorrected. 

The performance of any photometry tool is tied to fundamental constraints on the information that can be extracted from astronomical images. The underlying parameters $\theta$ that describe the sky signal, the raw image data $\mathcal{I}$, and processed data or downstream summary statistics $g(\mathcal{I})$ form a Markov chain $\theta \rightarrow \mathcal{I} \rightarrow g(\mathcal{I})$. As such, the data processing inequality requires that the mutual information between $\theta$ and $g(\mathcal{I})$ is always less than or equal to that between $\theta$ and the original data, i.e., $I(\theta; g(\mathcal{I})) \leq I(\theta; \mathcal{I})$ \citep{dpi}. While in some cases (often under strict assumptions) it is possible to construct ``sufficient statistics" which satisfy $I(\theta;g(\mathcal{I})) = I(\theta;\mathcal{I})$, methods that can directly access the mutual information between $\theta$ and $\mathcal{I}$ will be crucial for extracting the full information content of increasingly rich datasets.

The work presented in this paper builds on \emph{probabilistic cataloging} \citep[PCAT,][]{BREWER, BrewerDonovan, Daylan18, Kashyap14}, a framework that combines transdimensional inference \citep{GREEN} with Bayesian hierarchical modeling to sample from a \emph{metamodel} (union of models with different dimensionality) consistent with observed astronomical data. We extend the forward model to handle map data in which the observed signal can be composed as the sum of point sources, a diffuse fluctuation component modeled through a set of Fourier component templates, and surface brightness templates of unknown amplitude. This extension is implemented in the code \texttt{P}robabilistic \texttt{CAT}aloging in the presence of \texttt{D}iffuse 
\texttt{E}mission \citep[\texttt{PCAT-DE},][]{pcat_de_software}.

\pcatde\ is tested on a variety of synthetic observations from the Spectral and Photometric Imaging REceiver (SPIRE), an instrument on board the 3.5-meter \textit{Herschel} space observatory \citep{Griffin10}. The different applications in this work make assumptions about the spatial and spectral behavior of the components, however \pcatde\ is flexible and can handle the properties of different models as long as they are properly specified. Possible use cases include but are not limited to: separation of infrared sources and the CMB at sub-mm wavelengths; point-like source cataloging in the presence of large fixed-pattern detector noise; extraction of point sources over large-scale gradients caused by Zodiacal Light or fluctuations in atmospheric transmission/brightness for ground-based data; separation of X-ray point sources from diffuse galaxy cluster emission, among others. 

The paper is structured as follows. We begin in \S 2 with an introduction to probabilistic cataloging and its extension to model diffuse emission. The mock \emph{Herschel}-SPIRE data sets are introduced in \S 3 and we test the performance of our implementation on reconstructing blended emission components based on a range of models and data in \S 4. The first application explores how well point sources and their properties are detected/measured (\S 5), while the second models out the impact of point sources and diffuse emission on the thermal Sunyaev-Zel'dovich (SZ) effect (\S 6). We conclude in \S 7 with a discussion of the current \pcatde\ implementation and propose a number of potential applications for this formalism.

\section{Probabilistic cataloging}
As telescopes become more sensitive, source extraction becomes increasingly limited by the ability to spatially resolve overlapping sources \citep{LSST}. This is driven by the gap between flux sensitivity and angular resolution, which becomes important as one pushes to fainter depths. For current and near-future surveys, an increasingly large fraction of sources that would be reliably measured in isolation will be observed as partial or full blends with adjacent sources, complicating both the identification and measurement of bright and faint objects \citep{lsst22}. For some datasets a fast mapping rate is prioritized over angular resolution, and these surveys in particular will approach sensitivities where source blending is relevant, both in the spatial and spectral domains\footnote{For certain large area surveys the conventional catalog of sub-threshold point sources may not contain more information than the intensity maps themselves \citep{Schaan21, Cheng19}}. For example, source blending will be a major source of systematic uncertainty in a variety of Stage-IV cosmology surveys which rely on accurate galaxy photometry \citep{Melchior21}. 

These challenges motivate probabilistic cataloging. By sampling the full catalog space consistent with a given dataset, probabilistic cataloging can be used to infer both the properties of astronomical sources present \emph{and} the number of sources itself, above some flux density threshold. PCAT models sources below conventional significance thresholds (i.e., $<5\sigma$), which enables detection of faint sources and less biased constraints of bright sources with faint neighbors. As a Bayesian hierarchical modeling framework, PCAT is capable of incorporating complex information into a self-consistent model of the signal and data generating process, assuming knowledge of the causal chain that leads to observed data. In this approach, marginalization over different parameters is performed by collecting fair draws from the posterior of the forward model given the data. Composable models like those used in probabilistic cataloging are easily interpretable by directly testing the addition or removal of components or by modifying model priors.

Applied to single-band optical images of the globular cluster M2 taken from the Sloan Digital Sky Survey (SDSS), probabilistic cataloging recovers sources with completeness one magnitude deeper than the crowded-field photometry tool \texttt{DAOPHOT} \citep{DAOPHOT}, along with a lower false discovery rate for brighter sources \citep{PORTILLO_17}. Performance by these metrics is further improved by extending the hierarchical model to multi-band data, in which case the maps are fit simultaneously \citep{Feder20, DAYLAN_1}.  

\subsection{Modeling astronomical images}

Let $\lambda_{ij}^b$ denote the surface brightness in pixel $(i,j)$ of band $b$. The model used to generate images within \texttt{PCAT-DE} can be written as a sum over point sources and diffuse signals:

\begin{equation}
\begin{multlined}
\lambda_{ij}^b = \mathcal{P}^b \circledast \Bigl[\sum_{n=1}^{\textrm{N}_{\textrm{src}}} S_{n}^b \delta(x_i-x_{n}^b,y_j-y_{n}^b) \; + \\ A_b^{\textrm{t}} \sum_{t=1}^{\textrm{N}_{temp}} \textrm{I}^{\textrm{t}}_b(x_i,y_j) \: + \: B_{ij}^b \Bigr].
\end{multlined}
\label{eq:gen_model}
\end{equation}
In this equation, $\mathcal{P}^b$ is the beam function kernel which convolves the signal measured in band $b$ by the point spread function. The signal is decomposed into a sum of point sources with flux densities $\lbrace S_n\rbrace_{n=1}^{\textrm{N}_{src}}$, $\textrm{N}
_{\textrm{temp}}$ templates for resolved components with known position/spatial structure encoded in surface brightness templates $\lbrace\textrm{I}_b^t\rbrace$ and amplitudes $\lbrace A_b^t\rbrace$ (see \S \ref{sec:szeffect}), and a generic term for additional diffuse signal $B_{ij}^b$. In images with negligible diffuse structure (or a small enough field of view), a simpler mean normalization in each band, \Bzero, may be sufficient. We use $B_{ij}^b$ to specify diffuse signals without \emph{a priori} spatial structure, for which a more flexible, non-parametric model is used in signal reconstruction (see \S \ref{sec:fc_modl}).

\begin{figure}
    \centering
    \includegraphics[width=\linewidth]{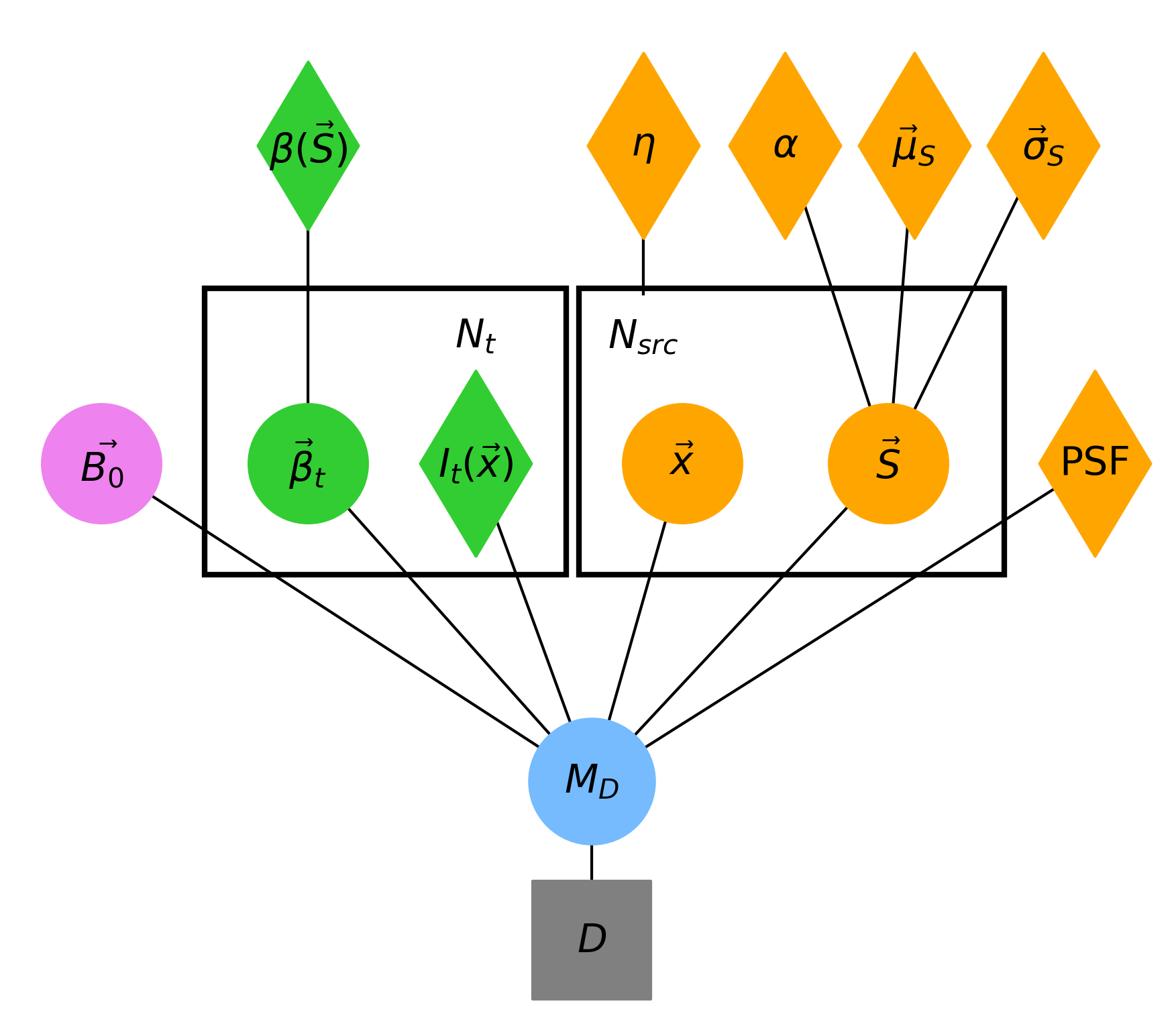}
    \caption{Probabilistic graphical model (PGM) for \texttt{PCAT-DE}. The top level of the PGM shows hyperparameters ($\alpha, \vec{\mu}_S, \vec{\sigma}_S, \eta, \beta(\vec{S})$) that characterize priors over the point source parameters $\lbrace \vec{x}, \vec{S}, N_{src} \rbrace$ and diffuse component colors $\lbrace\vec{\beta}_t\rbrace$. These parameters are then used to generate model images $M_D$ that are compared with the data. Diamonds and circles indicate variables that are fixed and floated, respectively.}
    \label{fig:bhm}
\end{figure}

For each band $b$, let $\vec{d}_b$ define the data vector corresponding to the unraveled image with size $\textrm{W}_b\times\textrm{H}_b$ pixels and corresponding per-pixel errors given by $\vec{\sigma}_b^2$. In this work we assume that errors are known for each pixel and independent of one another, meaning the likelihood can be written as a product over all pixels. We further assume these errors are Gaussian distributed. For the purpose of MCMC sampling, we compute log-likelihoods, turning the products over pixels and bands into sums:
\begin{equation}
\log \mathcal{L} \approx \sum_{b=1}^{\textrm{B}}\sum_{i=1}^{\textrm{W}_b}\sum_{j=1}^{\textrm{H}_b} - \frac{(\vec{d}_b-\vec{\lambda}_b)^2}{2\vec{\sigma}_b^2}.
\label{eq:logL}
\end{equation}
While the likelihood is in a space of fixed dimension set by the data, the space of models is transdimensional, i.e. it is the union of catalogs with varying N$_{src}$. Because these models reside in nested sub-spaces of one another, priors can be placed on individual mixture components (the point sources) while defining a posterior over the full model space \citep{PORTILLO_17}.

Figure \ref{fig:bhm} shows a representation of the \texttt{PCAT-DE} generative model as a probabilistic graphical model (PGM). The different layers of the PGM correspond to levels of the Bayesian hierarchical model -- at the highest level, priors on the point source population (a power-law flux prior for sources with spectral index $\alpha$ and Gaussian prior on colors with mean $\vec{\mu}_S$ and width $\vec{\sigma}_S$) and diffuse components (e.g., the color of the diffuse component encoded in $\beta(\vec{S})$) inform the prior distributions over mixture components. The point sources and diffuse components are then convolved by the instrument beam to produce model images that can be compared directly to the observed data. Note for this work that $N_t$ is fixed, while $N_{src}$ is floated as a free parameter.

\subsection{Modeling diffuse signals with Fourier component templates}\label{sec:fc_modl}
Diffuse signals are modeled by \pcatde\ using a linear combination of Fourier component templates, where each template represents a separate Fourier mode. An arbitrary diffuse signal can be approximated by a truncated two-dimensional Fourier series:
\begin{equation}
    B_{ij}^b = B_{0}^{b} + \sum_{n_x=1}^{\textrm{N}_{\textrm{FC}}}\sum_{n_y=1}^{\textrm{N}_{\textrm{FC}}}\pmb{\beta}_{n_x n_y}\cdot \pmb{\mathcal{F}}_{ij}^{n_x n_y}.
    \label{eq:totalbkg}
\end{equation}
Here \nfc\ refers to the order of the Fourier series and $\pmb{\mathcal{F}}_{ij}^{n_x n_y}$ is a vector of Fourier components corresponding to wavevector $(k_x, k_y) = (\textrm{W}/n_x,\textrm{H}/n_y)$ evaluated at pixel with index $(i,j)$:
\begin{equation}
    \pmb{\mathcal{F}}_{ij}^{n_x n_y} = \left(\begin{array}{c} \sin\left(\frac{n_x\pi x_j}{\textrm{W}}\right)\sin\left(\frac{n_y\pi y_j}{\textrm{H}}\right) \\ 
    \sin\left(\frac{n_x\pi x_j}{\textrm{W}}\right)\cos\left(\frac{n_y\pi y_j}{\textrm{H}}\right)
    \\
    \cos\left(\frac{n_x\pi x_j}{\textrm{W}}\right)\sin\left(\frac{n_y\pi y_j}{\textrm{H}}\right)
    \\
    \cos\left(\frac{n_x\pi x_j}{\textrm{W}}\right)\cos\left(\frac{n_y\pi y_j}{\textrm{H}}\right)
    \end{array}\right)
    \label{eq:fcomp}
\end{equation}
The vector $\pmb{\beta}_{n_x n_y}$ encodes the amplitudes of the Fourier components. All four components of  $\pmb{\mathcal{F}}_{ij}^{n_x n_y}$ are necessary in the absence of boundary conditions on the images, which in general will be arbitrary. Throughout this work we use the parameter \nfc\ when comparing models. The minimum angular scale captured by Fourier components can be approximated by the half-period of the highest frequency Fourier mode along each dimension\footnote{The smallest angular scale is actually set by the norm of the wavevector, $|k|_{max} = \sqrt{k_{x,max}^2 + k_{y,max}^2}$.}.


Fourier component templates are well suited for the tasks at hand. Using a truncated Fourier series has the benefit of robustness against bias from small-scale, localized signal map features (e.g., unmodeled point sources). This is because each template has global extent over the image and the choice of truncation scale implies certain modes are impossible to reconstruct with the Fourier series. For astronomical images, the minimum effective scale is typically bounded by the beam rather than the chosen map pixel size. In general the point source model provides a good description of fluctuations on the scale of the beam, however there is a range of intermediate scales larger than the beam and smaller than the image size where power from diffuse signals can reside. Fortunately, the falling angular power spectrum characteristic of many diffuse signals implies they may be well described by models for which $k_{\theta}^{max} < k_{\theta}^{beam}$, where $k_{\theta}^{beam}$ corresponds to the angular scale of the PSF full width at half maximum (FWHM). Many methods for point source detection, such as \texttt{SExtractor} \citep{sextractor} and \texttt{Starfinder} \citep{starfinder}, include empirical estimates of the local sky signal surrounding each source using a pixel-wise mean or median. The Fourier component representation is flexible enough to accomplish effective sky subtraction, however the underlying feature of PCAT's forward modeling is the capability to incorporate physically motivated priors for diffuse components within a larger Bayesian hierarchical model (including point sources and realistic observational noise).

A set of linear marginalization steps completed at the beginning of sampling (see Appendix \ref{sec:fc_marg}) accelerates the burn-in phase of sampling, after which the Fourier coefficients are sampled with the same Metropolis-Hastings algorithm used for the rest of the model parameters. In practice the Fourier components converge in a similar number of iterations as the rest of the model. The proposal kernel of each template is chosen by approximating the Fisher information of a uniform background component in the presence of several point sources (see Appendix \ref{sec:cov_ptsrc_bkg} for a derivation).

\section{Mock data}
\label{sec:mock_data}
In this section we describe the astrophysical components that are combined to generate mock observations  with similar noise properties as a range of shallow and deep SPIRE observations. SPIRE included a three-band imager with bandpasses centered at 250, 350 and 500 $\mu$m and beam FWHMs of 18\arcsec, 25\arcsec\ and 36\arcsec, respectively \citep{Griffin10}. While \pcatde\ has been applied to real SPIRE data in \cite{Butler21}, controlled sets of mocks are used here in order to characterize performance of the implementation in different limits. SPIRE maps typically contain a combination of emission from CIB galaxies and diffuse galactic cirrus. Galaxy cluster observations with SPIRE also contain localized but faint and extended signals from the thermal SZ effect. These maps are typically dominated by fluctuations in the total signal from individually undetected (and spatially unresolved) CIB galaxies, known as ``confusion noise"  \citep{Nguyen_2010, condon74}, providing a difficult scenario for point source extraction in the presence of diffuse contaminants. At the angular scale of the SPIRE beam, the underlying CIB luminosity function is extremely steep \citep[for differential SPIRE number counts see Figure 13 of][]{casey14}, resulting in a large number of sources just below the typical detection limit within a PSF-sized aperture. As a result, the source confusion in SPIRE observations should be considered more severe than the ``typical" use case in which sources are well separated and Poisson fluctuations are larger.

\subsection{Cosmic infrared background galaxies}
The majority of sources detected at far-infrared (FIR) wavelengths are $z\sim 2$ galaxies with a angular extent of $\sim$1\arcsec. When convolved with the much larger SPIRE beam, most galaxies in SPIRE observations are well modeled as point sources. Mock realizations are generated using the CIB model described in \citet[][referred to as \texttt{B12} throughout this work]{Bethermin2012}. On the scales of the images considered ($\theta \leq 10$ arcmin), the CIB power spectrum is dominated by shot noise from galaxies. More details about the construction of this CIB component can be found in \cite{Butler21}. 

\subsection{Galactic cirrus foregrounds}

A significant source of diffuse emission, even at high galactic latitudes, is Galactic cirrus dust, which reprocesses the interstellar radiation field and emits thermal radiation in the far-infrared \citep{cirrus}. While cirrus has a blue spectrum across the SPIRE bands similar to that of many observed CIB sources, cirrus contains most fluctuation power over larger angular scales. To calibrate the level of cirrus emission present in extragalactic observations for this study, we apply the \planck\ SZ-union foreground mask\footnote{Maps can be found here: \url{https://irsa.ipac.caltech.edu/data/Planck/release_1/ancillary-data/previews/COM_PCCS_SZ-unionMask_2048_R1.11/index.html}} to \planck\ observations, and sample positions uniformly across the unmasked sky. The maps at these positions are queried, re-gridded to SPIRE resolution and extrapolated to 250, 350 and 500 $\mu$m using the \planck-estimated parameters of a modified blackbody SED
\begin{equation}
    S(\nu) = A\left(\frac{\nu}{\nu_0}\right)^{\beta}B_{\nu}(T_d)
\end{equation}
where $\nu$ is the rest-frame frequency, $\nu_0$ is the reference frequency at which the optical depth is measured, $\beta$ is the spectral index, $T_d$ is the dust temperature and $B_{\nu}(T_d)$ is the spectral radiance at frequency $\nu$ assuming thermal equilibrium for temperature $T_d$. The  have dust temperatures ranging from 19-22K over 15\arcmin\ patches of the sky, while $\beta \sim 1.5\pm 0.1$ at the same smoothing scale \citep{planck_XI}.
We define a nominal ``1x-\planck" case as a diffuse signal whose power spectrum is parameterized by a single power-law slope:
\begin{equation}
    P(k) = P_0(|k|/k_0)^{\gamma}
\end{equation}
where $\gamma=-2.6$ \citep{Bracco11} and $P_0$ is determined by the unmasked sky-averaged power spectrum. Synthetic cirrus maps are drawn as Gaussian random realizations of the 1x-\planck\ power spectrum, and progressively more severe cirrus realizations are obtained by scaling the amplitude of fluctuations, $\sqrt{P(k_{\theta})}$, by factors of 2, 4 and 8. The range of cirrus realizations considered in this work with increasing fluctuation power are representative of the worst 50, 32, 17, and 5 per cent of the \planck\ unmasked sky. 

Lastly, mock observations are generated for a range of noise levels ranging from 1 \mjybm\ (confusion-dominated for SPIRE) to 6 \mjybm\ (instrument noise roughly equal to confusion noise). Instrument noise at the fiducial SPIRE map resolution is well described by a diagonal map-space covariance matrix \citep{Viero13}, and is dominated by thermal emission from the primary mirror.

\begin{figure*}
    \centering
    \includegraphics[width=0.95\linewidth]{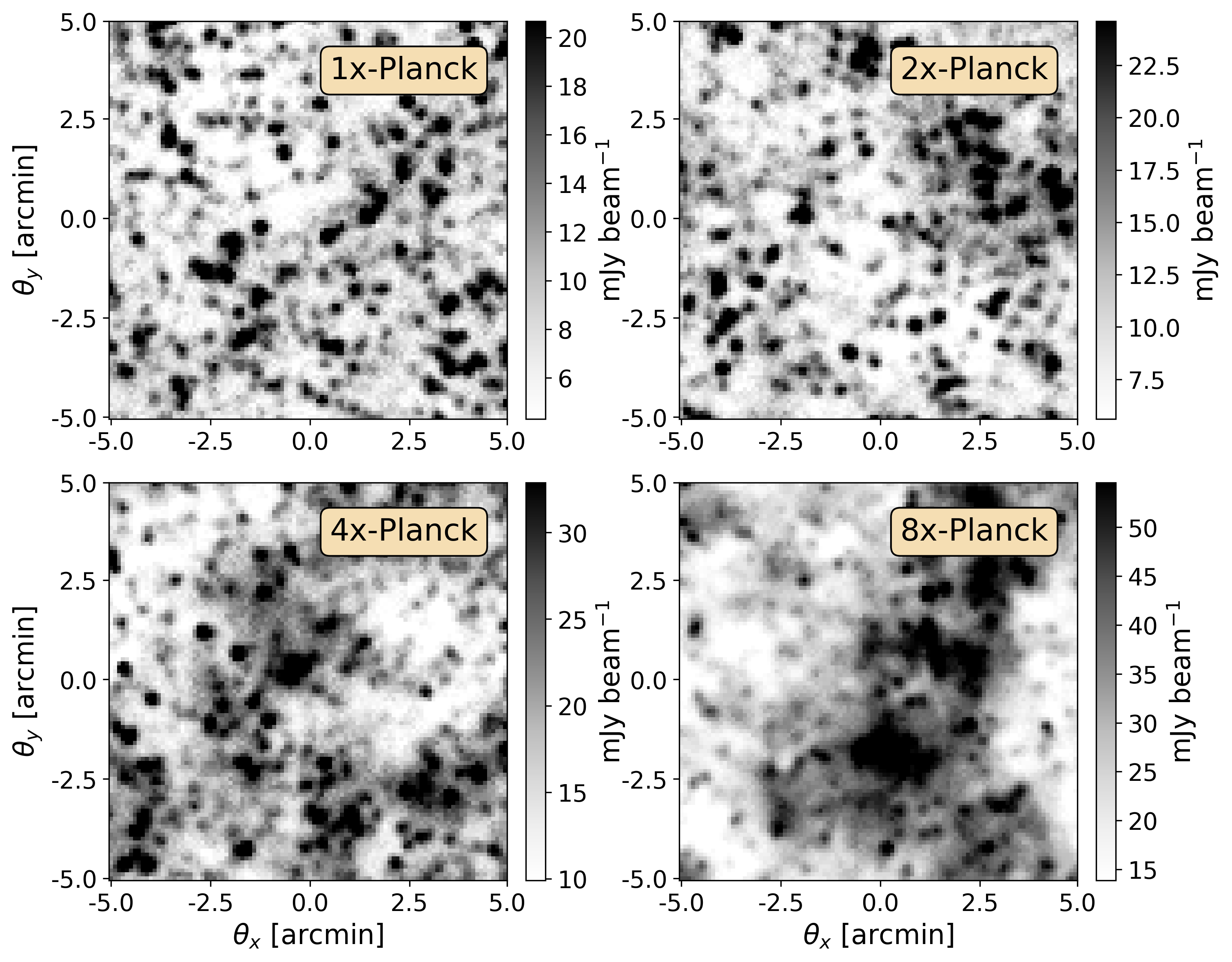}
    \caption{Mock realizations of CIB and cirrus at SPIRE resolution with 1 mJy beam$^{-1}$ instrument noise. In this limit confused point sources are the dominant agent reducing our modeling precision. Different panels show levels of cirrus consistent with the median ``clean sky" \planck\ sample (top left) and progressively more contaminated fields.}
    \label{fig:mock_cib_cirrus}
\end{figure*}

Figure \ref{fig:mock_cib_cirrus} shows a set of $10\arcmin \times 10\arcmin$ mock observations at 250 $\mu$m with varying levels of synthetic cirrus emission and instrument noise at the 1 \mjybm\ level. Uncorrected diffuse signals have the effect of boosting sources spatially coincident with positive fluctuations while suppressing sources coincident with negative fluctuations. Twenty sets of multiband CIB sky realizations are combined with different levels of synthetic cirrus throughout the results. In Sections 4 and 5 only the single-band 250 $\mu$m maps are used, however the full three-band maps are used in \S \ref{sec:szeffect}, where color information helps to distinguish the SZ effect from other astrophysical components.

\section{Separating point-like and diffuse emission in astronomical images}

\subsection{Model priors}
The priors used in this work are nearly identical to those from \cite{Butler21} and we summarize them briefly here.
A single power law flux distribution is assumed with slope $\alpha = -3.0$, and a uniformly distributed prior over the map is placed for source positions. A mininum flux density is imposed on the primary 250 $\mu$m band, set to $S_{min}^{250}=3$ and $S_{min}^{250}=5$ mJy for low/high instrument noise configurations respectively. We find the  results in this work are relatively insensitive to specific choice of \smin, though for rigorous characterization of sub-threshold number counts \smin\ can be varied as a hyperparameter. The mean additive normalization of each SPIRE map, denoted $B_0$, has no physical significance (i.e., SPIRE is not absolutely calibrated) and so we place a uniform prior on this component. Likewise we place uninformative priors on the amplitudes of the Fourier component template amplitudes, though a power spectrum prior is used in the initial set of Fourier component marginalization steps taken during burn in (see Appendix \ref{sec:fc_marg}).

One difference made in this work is the choice of prior on the number of sources, $\pi$(\nsrc), which counteracts the effect of overfitting due to additional parameters. As explained in Section 3.1 of \cite{DAYLAN_1}, PCAT uses two pairs of transdimensional MCMC proposals to explore the full catalog space. In the first pair a number of sources are chosen to either add (``births") with fluxes drawn from a flux+color prior, or remove sources at random (``deaths"). The second pair of proposals either split individual model sources into several or merge pairs of sources. As the number of degrees of freedom approaches infinity, the expected improvement in the log-likelihood approaches 1/2 for each additional parameter \citep{wilks}, implying an exponential prior on the number of sources
\begin{equation}
\label{eq:parsimony_prior}
    \log \frac{\pi(N+1)}{\pi(N)} = -\frac{1}{2}\rho\left(2+n_f\right).
\end{equation}
As the number of degrees of freedom approaches infinity, the average improvement in the log-likelihood approaches 1/2 per degree of freedom \citep{wilks}. For heavily confused observations the number of model parameters becomes non-negligible compared to the dimension of the data, which has the effect of increasing the amount of overfitting ($\langle \Delta \log \mathcal{L}\rangle > 0.5$ per d.o.f.). We use the scaling parameter $\rho$ to modify the parsimony prior. As such, $\rho$ may be derived \emph{a priori} by computing the ratio of $\langle \Delta \log \mathcal{L}\rangle$ for some fixed source number density above $S_{min}$ (plus any additional model parameters) with the same expectation in the ``sparse" limit. We derive an expression for $\langle \Delta \log \mathcal{L}\rangle$ in the non-asymptotic limit in Appendix \ref{sec:parsimony_prior_app}. For our runs we use $\rho = 1.5$, which is slightly more aggressive compared to that assuming single band source number densities from \cite{Bethermin2012} and $S_{min}=3$ mJy, which suggests $\rho \sim 1.35$ (see Fig. \ref{fig:parsimony_prior}).


The chains used throughout the paper were run for 4000 thinned samples each, where within \pcatde\ one thinned sample $= 10^3$ samples. We confirm that the chains converge within the first 2000 thinned samples, and we use the last 1000 samples from each chain for our results. 

\subsection{Number of Fourier components}
\label{sec:howmany}
The order of the Fourier component model is a hyperparameter that is chosen before running \pcatde. Including too few Fourier components may lead to residual diffuse fluctuation signal, however including too many components becomes computationally inefficient and makes the model more susceptible to overfitting. In principle, a fully transdimensional approach might infer the effective order of the Fourier component model. However, constructing efficient proposals that sample across Fourier component models of varying order is non-trivial because the number of parameters in the Fourier component model scales as  $k_{max}^2$, meaning a penalization based on the number of additional parameters becomes prohibitive. 

The hyperparameter \nfc\ is optimized by fitting several Fourier component models to mock data and comparing summary statistics as a function of \nfc. In general, we find that setting \nfc\ such that the highest angular frequency Fourier component has $k_{\theta}^{max}$ that is \emph{twice the cirrus-CIB power spectrum cross-over scale} (i.e., $k_{\theta}^{max}=2k_{\theta}^{\prime}$ where $P(k_{\theta}^{\prime})_{diffuse}=P(k_{\theta}^{\prime})_{shot}$) leads to an effective, parsimonious model, in the sense that the recovered diagnostics do not improve significantly by going to higher \nfc. A range of Fourier component models are tested ranging from \nfc=2 to \nfc=15, which correspond to truncation scales $\theta^{FC}_{min}$ of 5\arcmin\ and 40\arcsec\ arcseconds, respectively. For each cirrus amplitude case, the CIB realization is fixed to isolate trends due to varying \nfc.

\begin{figure}[h]
    \centering
    \includegraphics[width=0.95\linewidth]{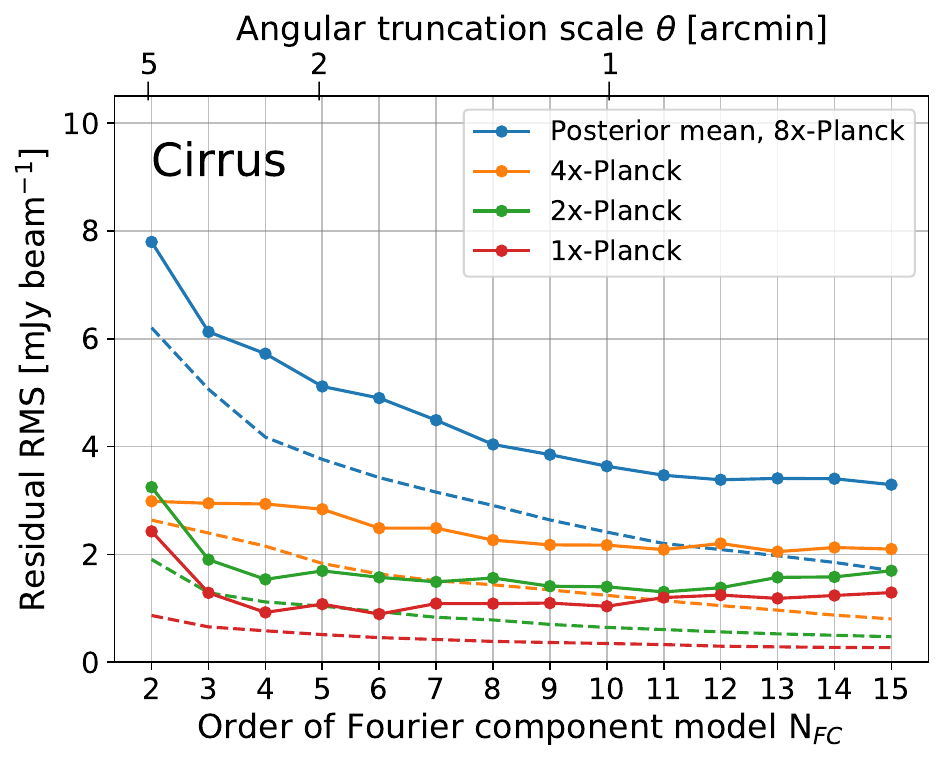}
    \includegraphics[width=0.95\linewidth]{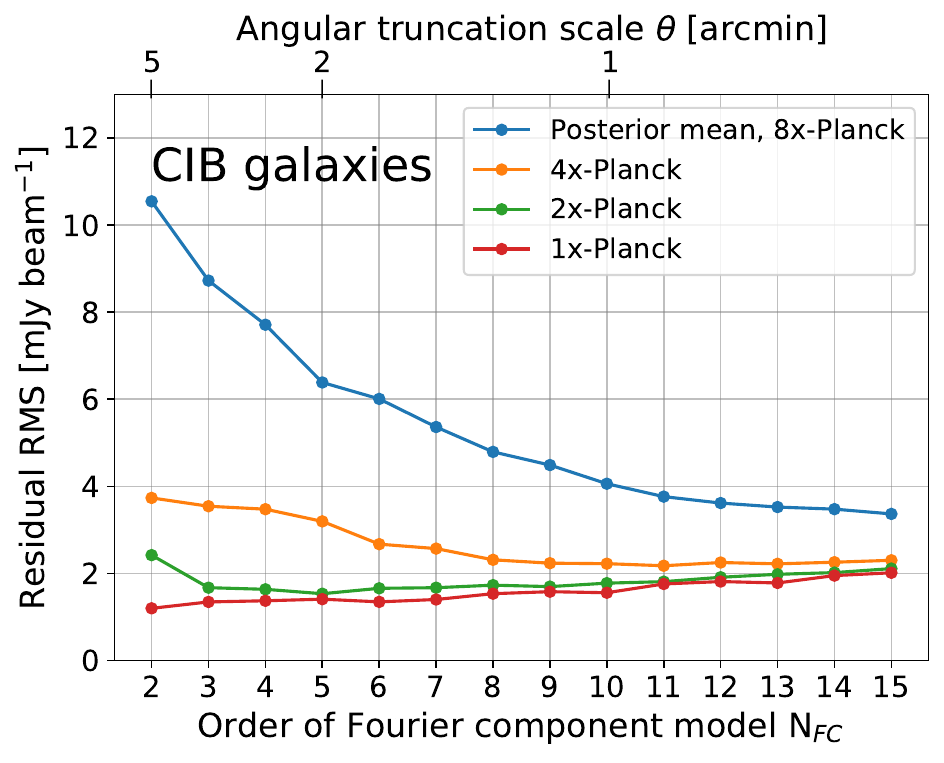}
    \caption{Reconstruction accuracy for diffuse cirrus (top) and CIB galaxies (bottom) in mock SPIRE observations, as a function of Fourier component truncation scale (ranging from $\theta^{min}_{FC}=5\arcmin$ to 40\arcsec). Different colors show how results change upon increasing the level of cirrus signal. The top axes indicate the approximate angular truncation scale corresponding to different \nfc. The dashed lines in the top figure show the residual RMS levels obtained from fitting Fourier components to the same cirrus realizations but without CIB injected.}
    \label{fig:rms_vs_nfc}
\end{figure}

Figure \ref{fig:rms_vs_nfc} shows the residual root-mean-square (RMS) averaged over pixels for both the recovered cirrus and CIB components. This statistic speaks to the general model reconstruction performance and how it changes with Fourier order, and by inspecting component-wise residuals we can assess the point beyond which additional Fourier components do not improve model performance. 
Within statistical errors, the cirrus residual level converges as the order of the Fourier component model increases. While the RMS is an incomplete measure for how well the data can constrain diffuse signals with arbitrary Fourier structure, the relative RMS contribution from degeneracies with point sources can be estimated. In particular, the MAP solution from each set of Fourier component templates is computed with respect to the same cirrus realizations, including instrument noise but without injected CIB. The errors from this simplified configuration are shown with the dashed lines in the top panel of Fig. \ref{fig:rms_vs_nfc}. These results suggest the RMS error due to the model fit quality and instrument noise is subdominant to confusion noise for CIB-dominated maps, while for maps with more fluctuation power from cirrus (e.g., 4x- and 8x-\planck), the error from each component is roughly equal.

Unlike the cirrus reconstruction which plateaus at larger \nfc, the CIB residual RMS levels increase by 60\% and 20\% relative to the minima of the 1x- and 2x-\planck\ cases, respectively. The goodness of fit does not change significantly across the same range, suggesting the increased component residual RMS is not due to an overall lack of convergence. A large portion of the CIB is undetected due to steeply falling number counts, so in the presence of a parsimony prior on point sources and absence of a power spectrum prior on the diffuse model, the Fourier components can preferentially absorb fluctuations from the CIB signal.

\begin{figure}
    \centering
    \includegraphics[width=\linewidth]{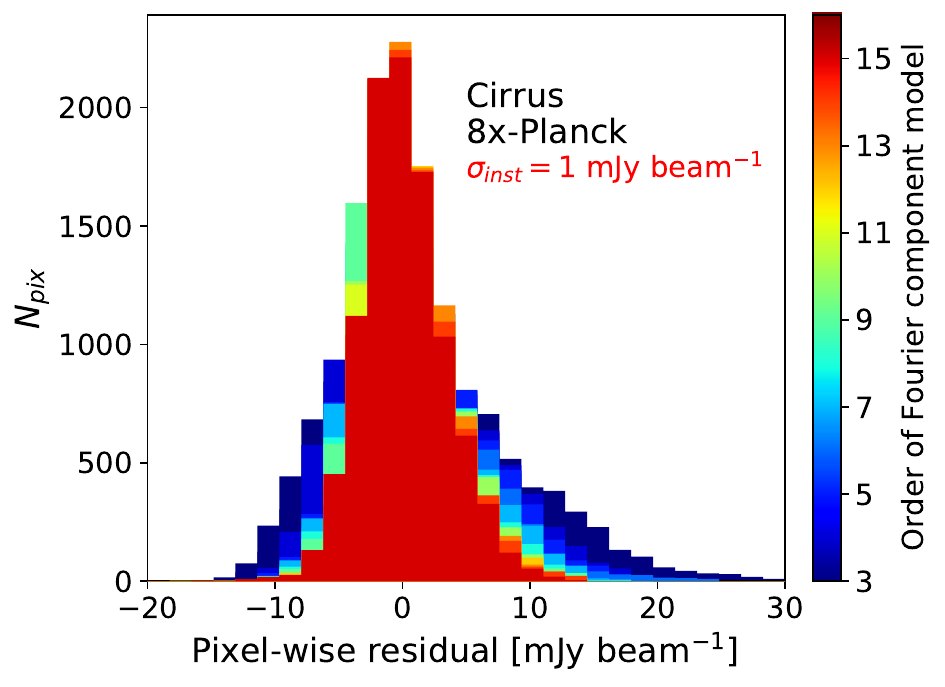}
    \includegraphics[width=\linewidth]{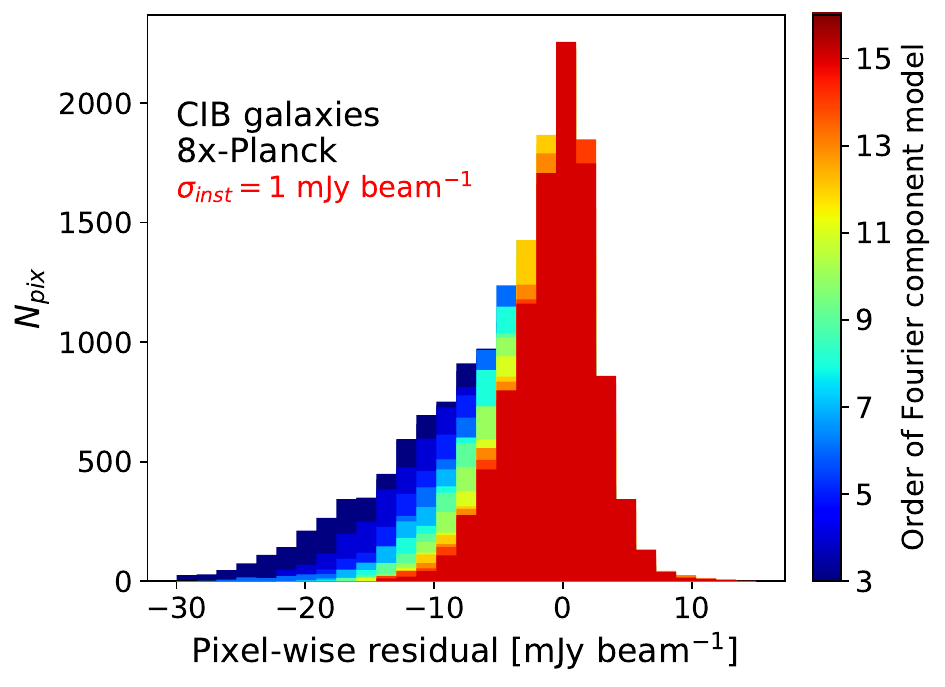}
    \caption{Histograms of pixel-wise residuals between cirrus (top), CIB galaxies (bottom) and the posterior mean of each respective model component.}
    \label{fig:comp_resid_hist}
\end{figure}

In addition to reducing RMS fluctuations in recovered signals, higher-order Fourier component models reduce skewness in the component-wise residuals. Figure \ref{fig:comp_resid_hist} shows the distribution of component-wise residuals over pixels for the most contaminated case (8x-\planck). As \nfc\ is increased from two to fifteen, the skewness in both CIB and cirrus residual is reduced considerably. The anticorrelation between CIB and cirrus residual 1-point distributions reflects oversubtraction of the CIB by spurious point sources, which may compensate for mismodeled diffuse signal when the order of the Fourier model is decreased.

For the remaining results, \nfc\ is fixed at each \planck\ cirrus level using the prescription described in \S \ref{sec:howmany}. This corresponds to \nfc\ = 5, 7, 11 and 15 for 1x-, 2x-, 4x- and 8x-\planck, respectively. 

\subsection{Component separation}
\begin{figure*}
    \centering
    \includegraphics[width=0.95\linewidth]{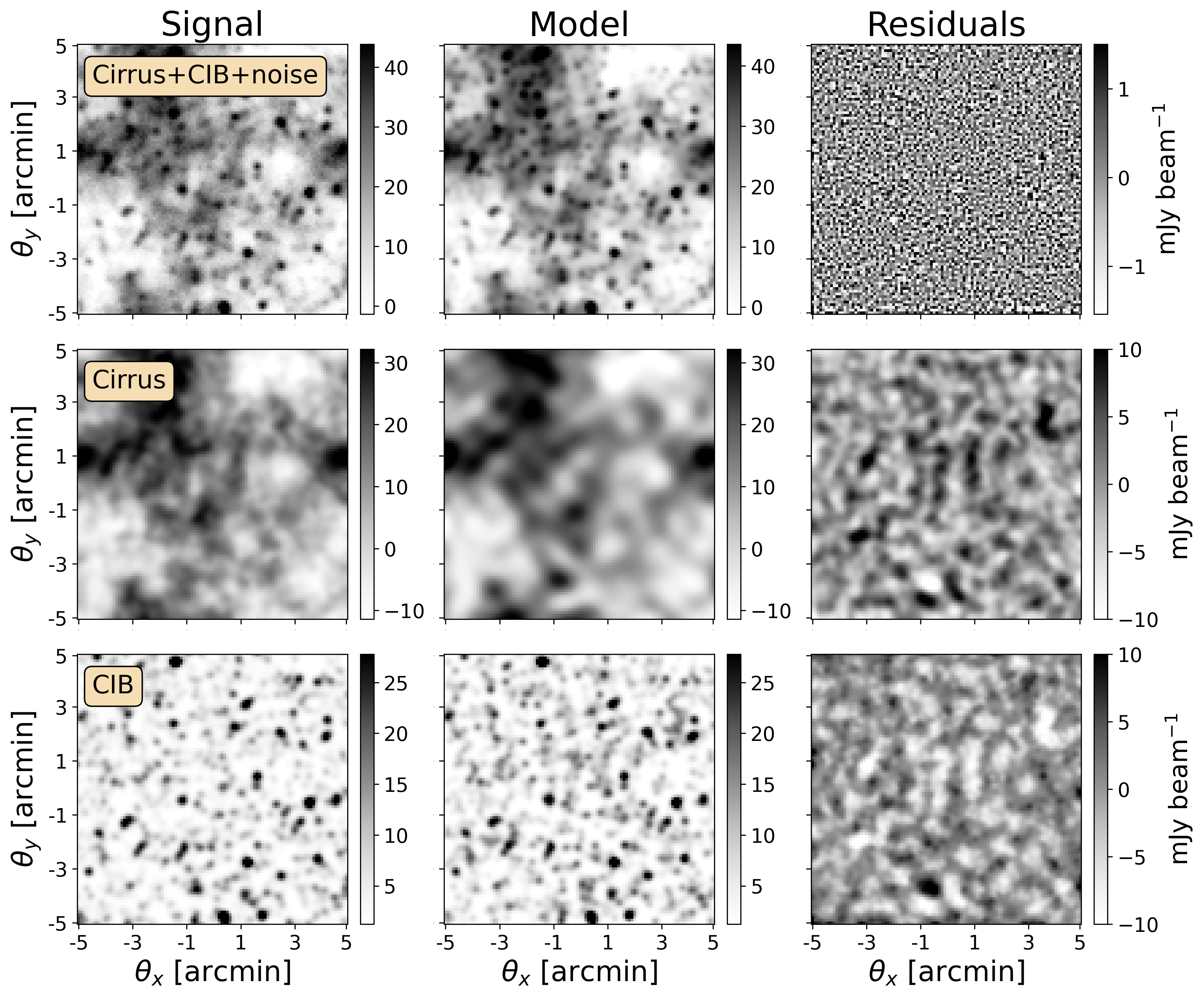}
    \caption{Component separation results for a mock CIB realization with injected cirrus dust (at the highest level, 8x-\planck) and SPIRE-like noise ($\sigma_{inst} = 1$ \mjybm). From the left, columns show the data signal (left), the median PCAT model (middle) and corresponding residuals (right).}
    \label{fig:imspace_resid_joint}
\end{figure*}

Figure \ref{fig:imspace_resid_joint} shows the input and recovered component maps for a mock CIB observation with cirrus at the 8x-\planck\ level, i.e. top 5\% of most contaminated \planck\ clean-sky observations. While the residual of the full model is consistent with noise, inspection of the individual component residuals shows errors in the recovered CIB and cirrus components. Notably, these errors are anti-correlated. The primary failure mode occurs when spurious sources compensate for spatially coincident residual diffuse signal, rather than from the diffuse signal concealing true point sources (this is supported by the results of \S \ref{sec:ptsrc}). Likewise, we find that the CIB residual is weakly correlated with the cirrus spatial curvature (Pearson correlation coefficient $\rho = +0.2$) and the positions of spurious point sources. Regions of high curvature imply presence of higher angular frequency modes that may be difficult to capture with the truncated Fourier series, so this correlation is expected.

\subsection{Component-wise power spectrum recovery}
\begin{figure}
    \centering
    \includegraphics[width=\linewidth]{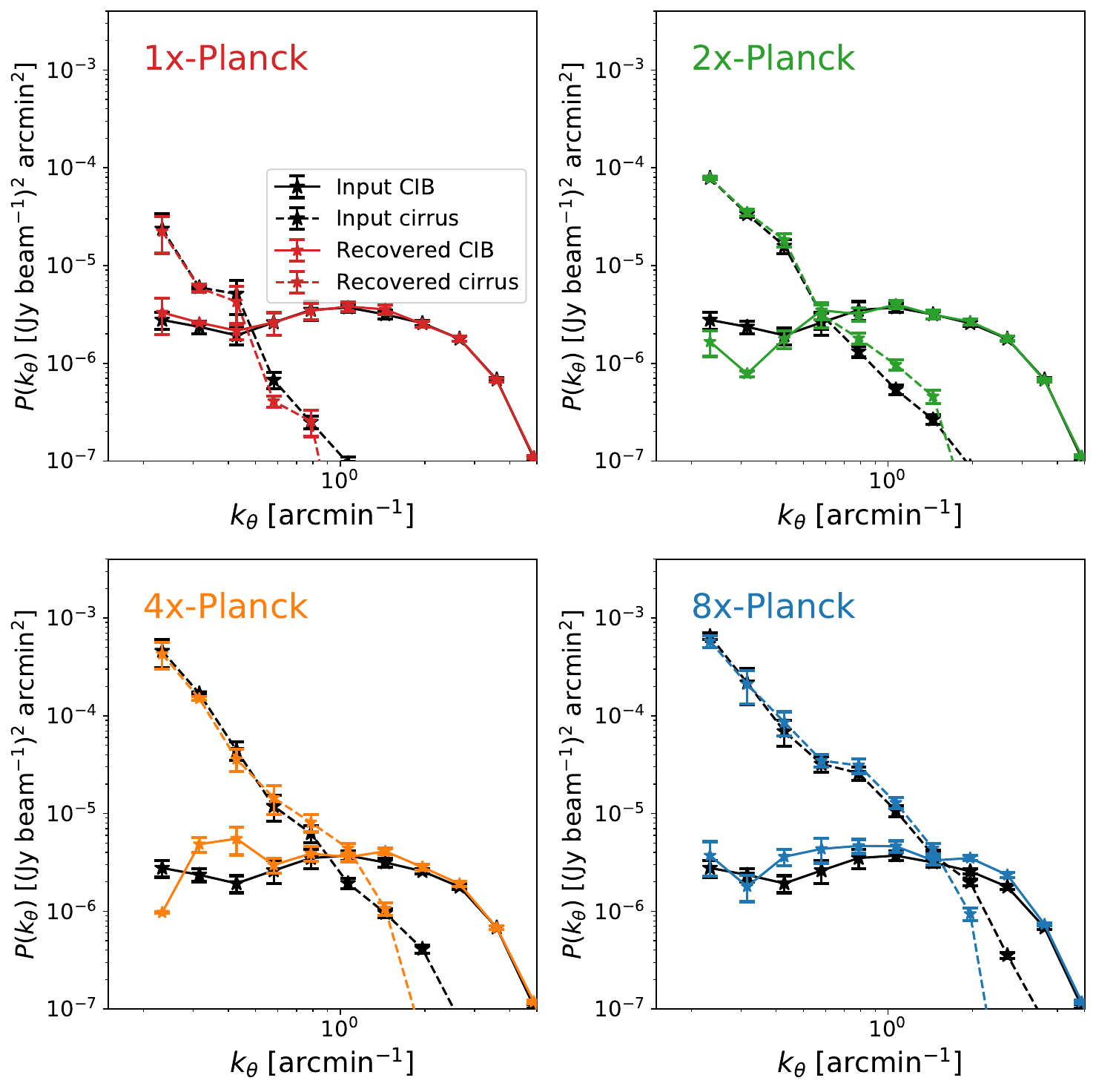}
    \caption{Comparison of input and recovered CIB (solid) and cirrus (dashed) power spectra for four cirrus realizations of increasing power. The recovered cirrus is represented by the Fourier component model image, while the CIB comes from the point source model. The per-pixel instrument noise is $10^{-3}$ Jy beam$^{-1}$, corresponding to a power spectrum amplitude of $10^{-8}$ Jy$^2$ beam$^{-2}$ arcmin$^2$.}
    \label{fig:powerspec_recover_comp_allx}
\end{figure}

The \pcatde\ model separates signals effectively in both map space and Fourier space. Figure \ref{fig:powerspec_recover_comp_allx} shows the recovered component power spectra of observations using a fixed CIB realization and four \planck\ cirrus realizations of increasing fluctuation power. The power spectra are computed from the model images with a Hanning window to mitigate spurious fluctuations sourced by the map boundaries. In nearly all cases, the power spectra of both components are reliably recovered, while for the 4x- and 8x-\planck\ cases the recovered CIB has a slight positive bias, which can be attributed to leakage from the much brighter cirrus signals. While cirrus-dominated observations have more false detections and faint-end flux boosting on average (cf. \S \ref{sec:ptsrc}), the residual fluctuation power of the CIB signal remains relatively small. This is reasonable in the limit where false detections are unclustered, i.e., they contribute to the mean normalization of the component model but not to its fluctuations. Low-level systematic biases in component separation with \texttt{PCAT-DE} may be more important to quantify in studies of the large-scale ($\theta>10\arcmin$) CIB clustering power spectrum, where linear clustering and diffuse emission are spatially degenerate.

\subsection{Computational requirements}
Forward modeling approaches like probabilistic cataloging are computationally demanding but tractable for targeted science fields. Proposals that perturb the template and mean normalization components are the dominant computational expense for \pcatde\ because they involve evaluating the delta log-likelihood over the full image or set of images, with an execution time that scales with the total number of pixels. For a fixed effective sample size (ESS), the time to obtain an independent sample from the chain naively scales as the square of the number of parameters when using Metropolis-Hastings proposals \citep{DAYLAN_1}. While this is mitigated for the point sources by evaluating the likelihood of point source proposals in smaller image patches, this is not possible with the Fourier templates which are defined over the full region of interest (ROI). In the absence of mean background and template-based proposals PCAT takes $\sim 30$ minutes in wall clock time to fit a 100$\times$100 pixel SPIRE image on a Macbook Pro with a 2.2 GHz Intel Core i7 processor using the Intel Math Kernel Library (MKL), and $\sim 1$ CPU hour without the MKL library. With mean background and template-based proposals included in the Metropolis-Hastings sampling, the wall clock time increases by a factor of $\sim 2$. 

Despite the computational challenge it should be possible to make the sampling algorithm more efficient. One option to speed up the performance is by executing marginalization steps, as are currently used during burn-in within \pcatde\ to accelerate $\chi^2$ minimization, intermittently during sampling, making \pcatde\ similar to a collapsed Gibbs sampler \citep{collapsed_gibbs}. The marginalization step integrates out the coefficients of the Fourier component model while fixing the remaining parameters at a given sample and may be more appropriate for analyses where the diffuse component coefficients are nuisance parameters. More generally, efficient proposals and sampling schemes can reduce the run time required for chain convergence and a sufficient ESS.

\section{Point source detection and population inference in the presence of diffuse emission}
\label{sec:ptsrc}
By incorporating a Fourier component model into source detection and deblending, probabilistic cataloging can recover sources obscured by negative diffuse signal fluctuations (relative to some mean normalization of the image) and reduce the number of false detections and boosted sources. In this section we test \pcatde\ on a set of CIB realizations with known positions/flux densities, from which we can examine the collection of detected sources and their properties as the level of injected cirrus is gradually increased. 

\begin{figure*}
    \centering
    \includegraphics[width=\linewidth]{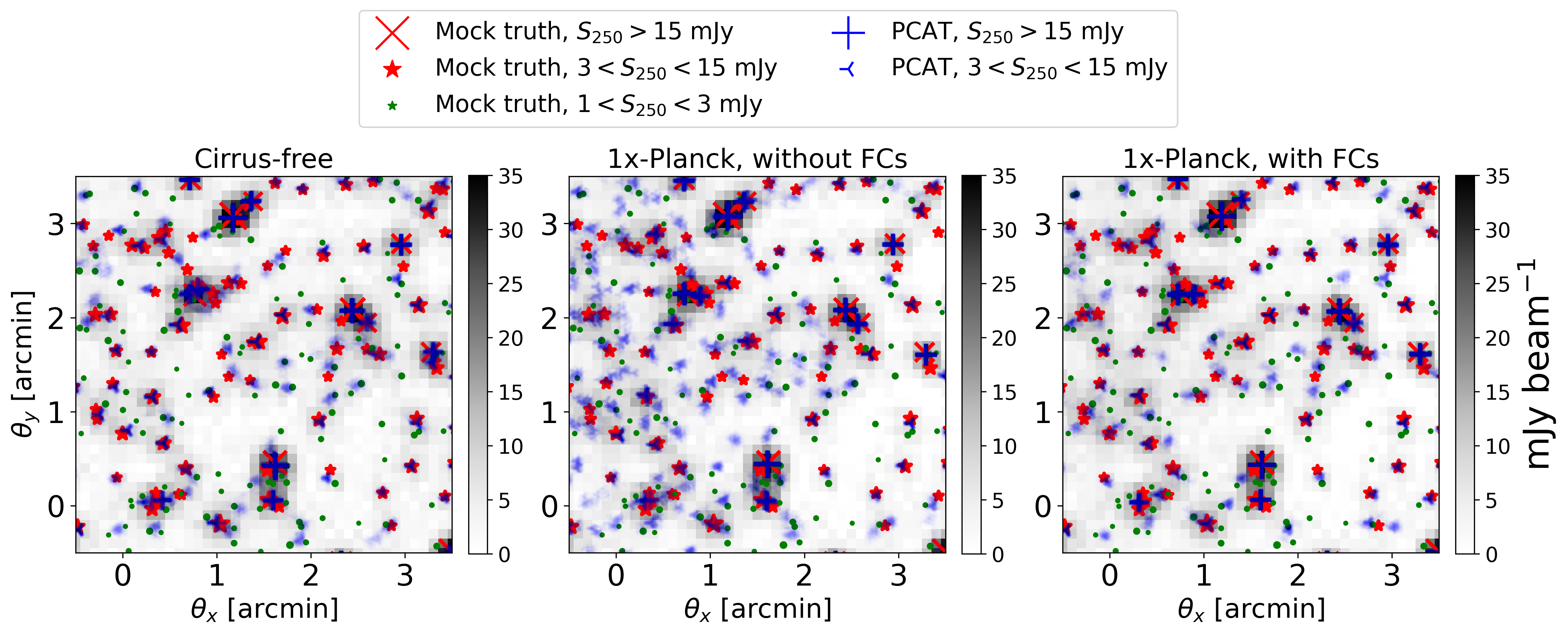}
    \caption{Comparison of recovered catalog ensembles for  a fixed 250 $\mu$m SPIRE CIB observation in the cirrus-free and 1x-\planck\ cases, with and without the use of Fourier components. In each panel, red symbols indicate the true positions of the input catalog down to $S_{250}=3$ mJy, while the blue shows 200 stacked samples that are uniformly distributed across the last 1000 posterior samples. ``Fuzzier" regions reflect the posterior uncertainty recovered in sources, typically on the faint end. The left panel shows PCAT run on CIB and instrument noise realizations, with no Fourier component model. The middle and right panels show results based on the same CIB realization but with additional synthetic cirrus dust drawn at the 1x-\planck\ cirrus level. While PCAT infers several spurious model sources in the absence of a diffuse signal model (middle), the inclusion of the Fourier component model leads to a recovered catalog ensemble nearly identical to the cirrus-free case (right).}
    \label{fig:with_without_fc}
\end{figure*}

Probabilistic cataloging requires precise control over systematic effects in observed data in order to constrain point source populations without incurring substantial errors. This is a consequence of the general fact that when a finite mixture model is misspecified (e.g., when it does not fully describe the data), the posterior on the number of components can diverge \citep{fmm}. Within probabilistic cataloging a minimum flux density is chosen for computational convenience, but also represents an instance of model misspecification, i.e., the true number counts extend below \smin. Diffuse signals are relevant in this context as well -- while they may not even be visible in an image, neglecting them when modeling observations can lead to biases on downstream measurements that rely on catalogs as starting points. These effects can be seen in Fig. \ref{fig:with_without_fc}, where catalog ensembles from \pcatde\ are compared with ground truth catalogs for three different runs. The middle panel shows how running PCAT on low-level, unmodeled cirrus leads to several spurious point sources clustered on the scale of the beam. The spurious sources are correlated with the gradient of the cirrus emission (along the $\theta_x$ direction), residing in regions where the mean normalization underestimates the diffuse component. The positions of spurious sources tend to trace faint underlying sources with flux densities that are below \smin\ (green points), suggesting that in this case small errors in the diffuse model primarily flux boost existing sources rather than generate completely fictitious sources. When the mean normalization overestimates the diffuse component, the modeled source flux densities will tend to bias low, which may also lead to a degradation of the catalog completeness for sources near \smin. The inclusion of a simple Fourier component model ameliorates the effects of foreground contamination significantly, with the recovered catalog posterior (right panel) nearly identical to the cirrus-free case (left panel). The stacked samples shown in blue can be converted into a ``condensed catalog" using an iterative cross-matching procedure \citep[see][for an outlined procedure]{PORTILLO_17}, with posteriors obtained from the collection of samples near each source. One can see visually in Fig. \ref{fig:with_without_fc} that brighter sources have more compact stacked samples, i.e., the posteriors on positions are well constrained. On the other hand, low significance sources and/or fictitious sources sourced by cirrus systematics are ``fuzzier", corresponding to posteriors that are much less constrained and which deviate from idealized Gaussian uncertainties\footnote{The departure from idealized uncertainties (assuming well isolated point source, perfect background subtraction) is quantified with the \emph{degradation factor} \citep[see Appendix C of][]{PORTILLO_17}. That work demonstrated that the degradation factors for positions and fluxes are highly correlated.}.

There is an intrinsic labeling degeneracy in probabilistic cataloging due to the fact that the number of sources is not fixed. As such, we compute metrics related to catalog completeness and reliability as expectations over the catalog ensemble returned by PCAT. We calculate the completeness of each catalog sample by finding the the closest model source to each true source within 6$\arcsec$ (one-third of SPIRE 250 $\mu$m beam FWHM) without replacement (i.e., the same PCAT source cannot be matched to several true catalog sources). Any PCAT source that has no true counterpart above \smin\ after this cross-matching procedure is classified as spurious. While a more stringent cross-matching procedure might include a match on flux density or log-fluxes, we are primarily interested in trends of these statistics with varying Fourier order. The level of blending for SPIRE sources further complicates interpretation of more detailed cross-matches (see \S \ref{sec:fluxboost_predict} for more details). 

\begin{figure}
    \centering
    \includegraphics[width=\linewidth]{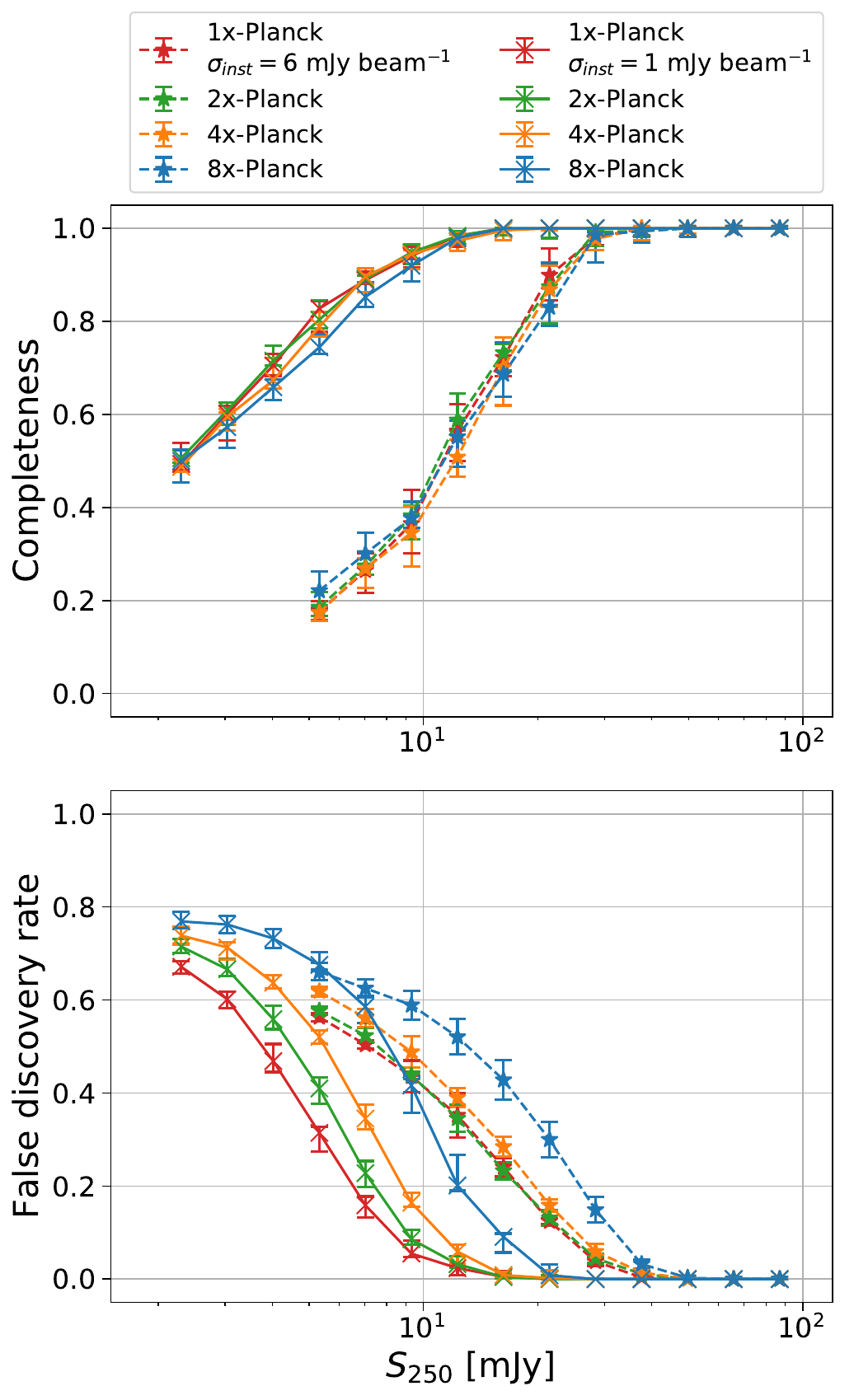}
    \caption{Completeness (top) and false discovery rate (bottom) of PCAT-detected sources for mock CIB data with $\sigma_{inst}=1$ \mjybm\ (solid) and $\sigma_{inst}=6$ \mjybm\ (dashed). Each line shows the mean and scatter from twenty mock CIB realizations.}
    \label{fig:comp_fdr_dust}
\end{figure}

Figure \ref{fig:comp_fdr_dust} shows the completeness and false discovery as a function of flux density, evaluated for our ensemble of CIB mocks at two noise levels (1 and 6 \mjybm). As the level of cirrus contamination increases, fainter sources become suppressed or entirely subsumed by diffuse signal fluctuations, leading to a mild degradation in source recovery. In contrast, the 90\% source reliability thresholds degrade from 8 (25) mJy for the 1x-\planck\ low- (high-) noise configurations to 16 (35) mJy for 8x-\planck. Spurious sources are included when the improvement in the log-likelihood from modeling residual diffuse emission with a spurious point source is greater than the penalty from adding parameters to the model and any other priors. The false discovery rate is also sensitive to the minimum source flux density permitted by the model.

\subsection{Predicting source blending}
\label{sec:fluxboost_predict}

Measuring the number density of sources as a function of flux density is a core task in astronomy. To recover correct number counts, one makes corrections for all effects that lead to observed counts, including catalog completeness, survey selections, astrophysical uncertainties, etc. One consideration is “flux boosting”, an Eddington-like bias where symmetric flux uncertainties added to sources drawn from a steeply falling luminosity function source an asymmetric scatter in the observed counts. In the context of sub-mm analyses, flux boosting can also be sourced by faint blended neighbors. Blending effects in single-dish sub-mm observations are often so severe that number counts are estimated from the one-point distribution of the maps (P(D) analysis, see \cite{pdanalysis_1, pdanalysis_2}) rather than from catalogs with individual sources. As a result, the method is limited to fields that are free of contamination from other components that would otherwise contribute to the skewness of the one-point function.

The flux boosting induced by source confusion can be well approximated through Bayesian model comparison. Using mock catalogs, we predict the “blended” catalog by identifying potential blends and evaluating the delta log-likelihood between two- and one-source models. This is an approximation to the full transdimensional inference performed with probabilistic cataloging. For a given blend of two (or more) sources, probabilistic cataloging estimates the relative Bayesian evidence between models with different N$_{src}$. If the likelihood does not improve significantly for observations with high underlying source multiplicity, PCAT will favor a simpler model to describe the observed signal. 


We predict the level of flux boosting for a given catalog in a probabilistic manner, evaluating the delta log-likelihood of two- and one-source models fit to underlying two-source configurations that might be blended by PCAT. This approach assumes prior information about the number counts of the underlying distribution, however marginalizing over uncertainties of the faint end LF is straightforward with this method if synthetic catalogs are available.

The delta log-likelihood for each candidate blend is calculated as follows: the best-fit one-source model position is approximated to be at the position where a PSF has the maximum overlap integral with the sum of two PSFs with positions and amplitudes corresponding to the two catalog sources. This approximation is exact in the infinite signal-to-noise limit. The maximum overlap integral position is on the line connecting the two sources and its distance along this line depends only on the two sources' flux ratio and separation. Then, we approximate the best-fit two-source model positions with the catalog positions and calculate the expected delta log-likelihood between the one-source model and the two-source model. Again, the best-fit two-source model positions are equal to the true (catalog) positions in the infinite signal-to-noise limit. While the maximum overlap position does not depend on the noise level, the expected delta-log likelihood does. Our validation of these two approximations using simulated images of pairs of point sources will be presented in a future manuscript. 

The delta log-likelihood $\Delta \log\mathcal{L}$ for a given pair of sources can be combined with the relative parsimony prior, $\log(\pi(N=2))-\log(\pi(N=1))$ (using equation \eqref{eq:parsimony_prior} with $\rho=1.5$) to obtain a delta log-posterior between models, $\Delta \log \mathcal{P} = \Delta \log\mathcal{L} + \Delta \log \pi$. For this calculation we ignore differences in posterior volume, though these differences are used in calculating acceptance probabilities within \pcatde. Let us assume that $P(N=2)+P(N=1) \approx 1$, i.e. $P(N>2) \ll 1$. Then $\Delta \log \mathcal{P}$ is related to the deblending probability by the following:
\begin{equation}
    p(N=2) \approx \frac{1}{1 + e^{-\Delta \log \mathcal{P}}}
\end{equation}
The algorithm iteratively evaluates blends, starting with the brightest source and finding the brightest neighbor within one FWHM of the source position. If there is no neighbor in the vicinity, the recovered flux is assumed to be the true flux, on average. If there is a neighbor, the delta log-likelihood of the two-source configuration is used to simulate blending by making a draw on a Bernoulli distribution with parameter $p(\textrm{N}=2)$. If the draw results in a blend, the best fit flux/position of the one-source model is added to the catalog and both original sources are removed. Once the full catalog has been processed in this way, the number counts are recomputed. These ``recovered" catalogs should more closely resemble the recovered flux distribution using a Bayesian approach like probabilistic cataloging. These recovered catalogs encode an approximation of the posterior distribution.
\begin{figure}
    \centering
    \includegraphics[width=\linewidth]{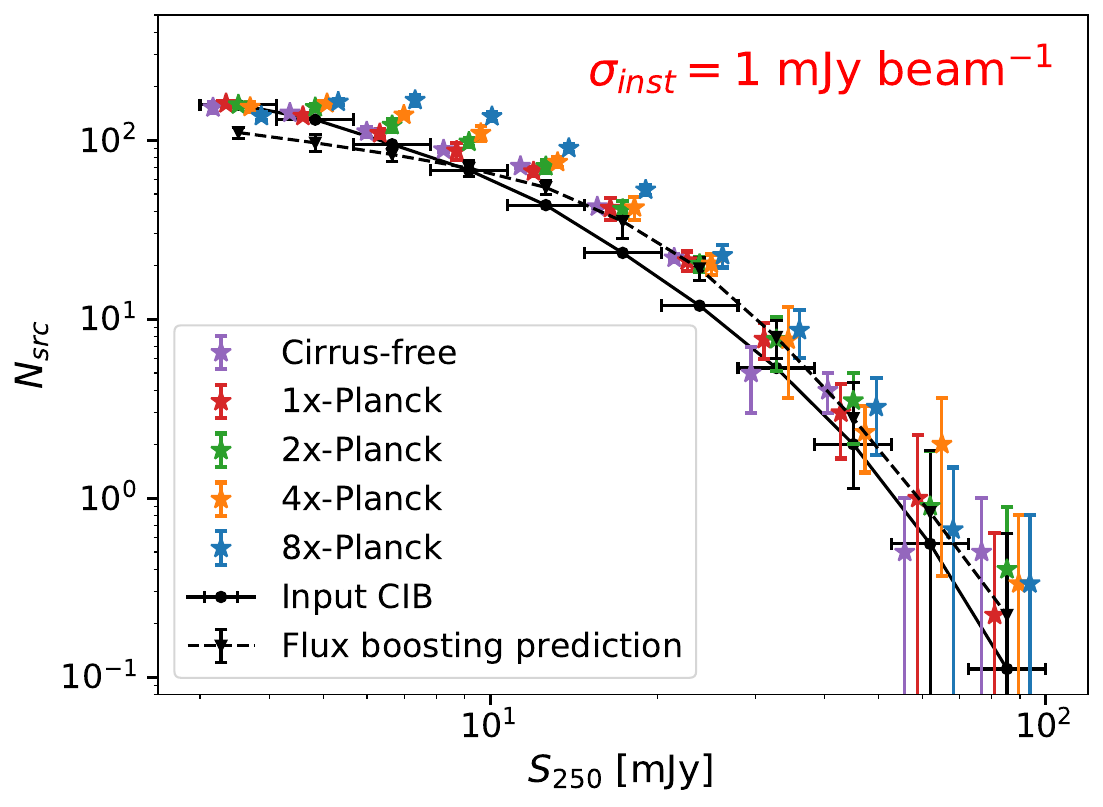}
    \caption{Input and recovered flux distributions for twenty mock CIB realizations with increasing cirrus contamination ($\sim 0.3$ deg$^2$ in total area). The recovered flux distributions are shifted slightly along the x-axis to highlight the dependence on diffuse contamination for each bin. Error bars indicate scatter across the $10\arcmin\times10\arcmin$ CIB realizations. The same set of catalogs is used as input for each cirrus level in order to reduce additional uncertainties due to sample variance.}
    \label{fig:flux_dist}
\end{figure}
Through our simulated blending procedure, we find the predicted number counts are consistent with those recovered using \pcatde\ for a range of flux densities.
Figure \ref{fig:flux_dist} shows the input and recovered flux distributions for different levels of cirrus contamination. All of the recovered flux distributions show overproduction of intermediate/bright sources relative to the input CIB catalog. Given the luminosity function of sub-millimeter galaxies and the angular resolution of SPIRE, this behavior is explainable by source blending. For a given blend of two (or more) sources, PCAT estimates the relative Bayesian evidence between models with different \nsrc. If the likelihood does not improve significantly for observations with high underlying source multiplicity, PCAT will favor a simpler model to describe the observed signal. This means for some range of source separations there is not enough information in the observed data to properly de-blend sources. This is a well known limitation for analyses of \emph{Herschel}-SPIRE data, and there are methods in the literature to correct for this mode of flux-boosting, both for individual objects and at the population level \citep{Coppin05, Crawford2010}. The number counts predictions including effects of blending closely match those obtained with \pcatde. The prediction does not take into account the covariance of \nsrc$>2$ source configurations, nor the prior volume effects associated with source parameters, and so they will be less accurate for fainter source flux densities. 

Nonetheless, Fig. \ref{fig:flux_dist} shows that on the bright end ($S_{250} > 20$ mJy) the recovered flux distributions are insensitive to all levels of injected cirrus. For $S_{250} \leq 20$ mJy the recovered number counts become increasingly correlated with the injected cirrus level. While the observed flux boosting of an analysis procedure will depend on the details of implementation, these results suggest it should be possible to empirically de-boost the observed number counts as a function of foreground contamination. 
\subsection{Sensitivity to \nfc}
\label{sec:ptsrc_nfc}
\begin{figure}
    \centering
    \includegraphics[width=\linewidth]{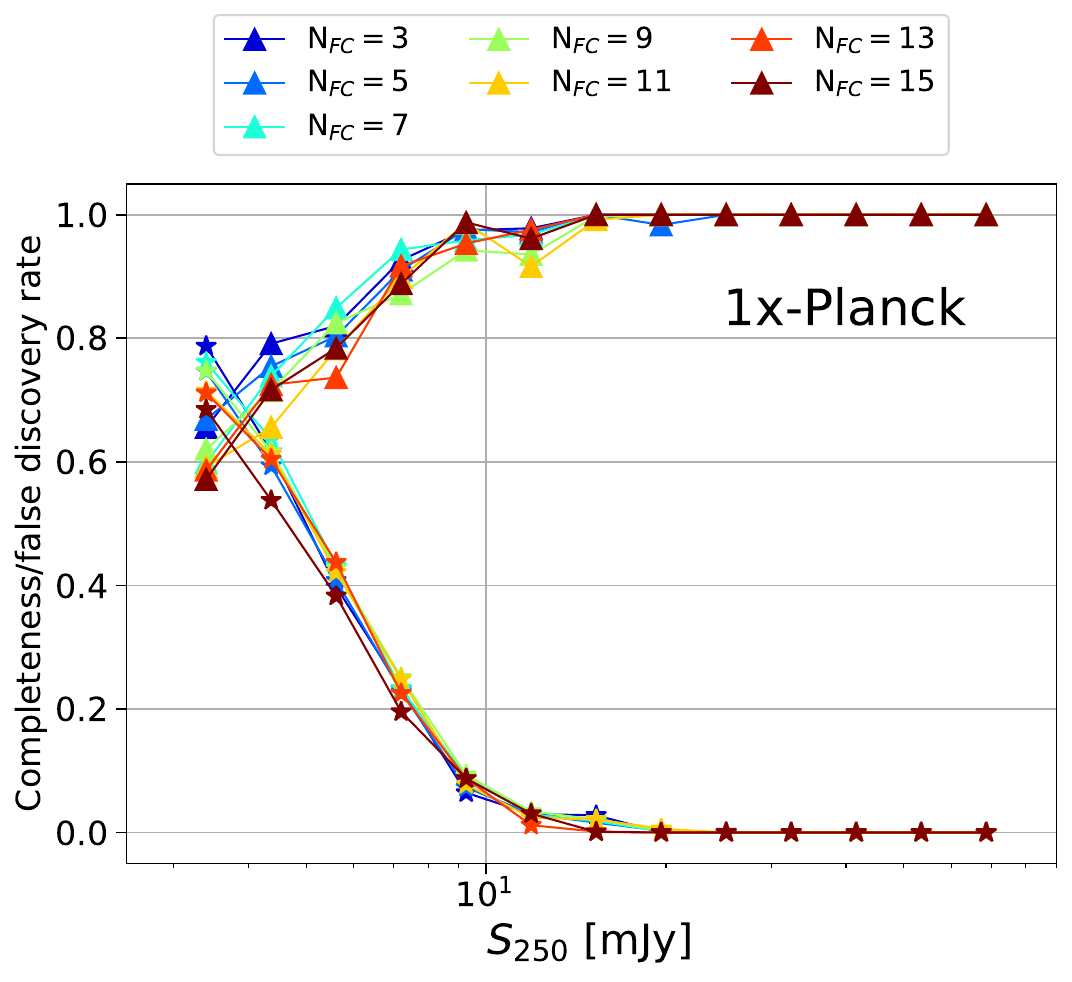}
    \includegraphics[width=\linewidth]{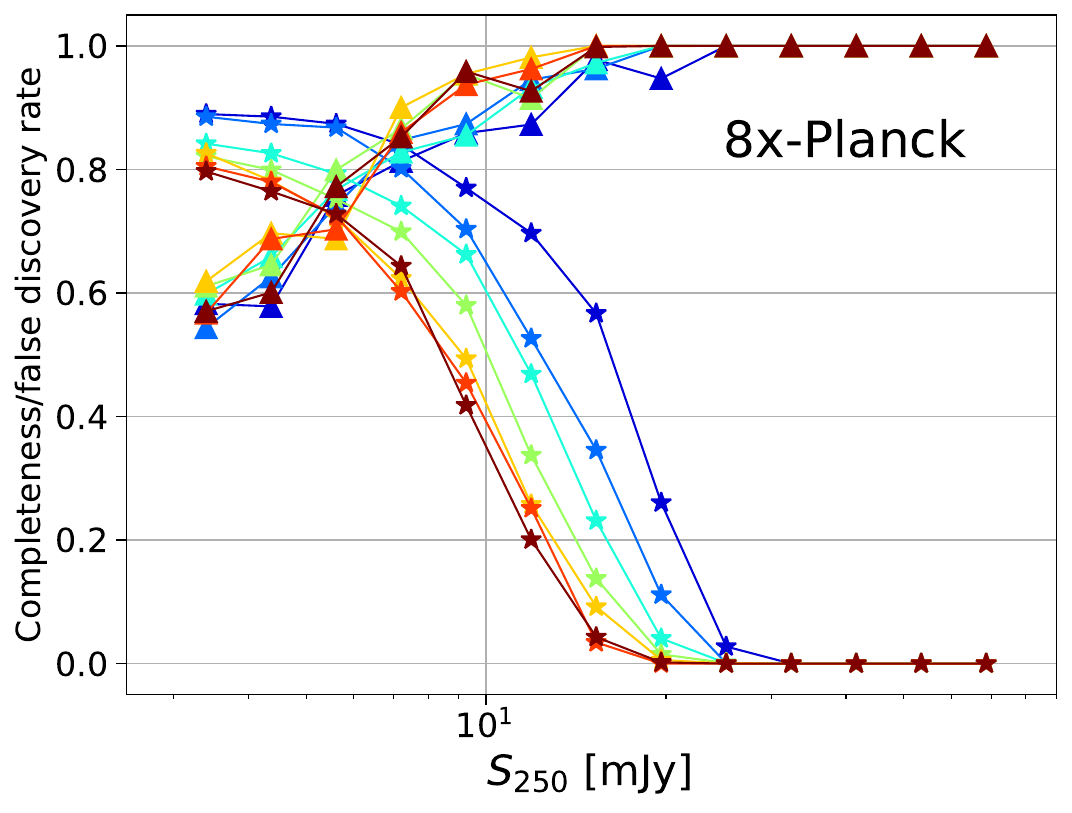}
    \caption{Completeness (triangles) and false discovery rates (stars) as a function of flux density. These are evaluated for a range of Fourier component models indicated by the different colors. The top panel shows the 1x-\planck\ results while the bottom shows the same for the 8x-\planck\ case.}
    \label{fig:comp_fdr_vs_nfc}
\end{figure}

Figure \ref{fig:comp_fdr_vs_nfc} shows the completeness and false discovery rate as a function of flux density for a subset of Fourier component models spanning the same range in \nfc\ as tested in \S 4.1. For the 1x-\planck\ mocks, there is little to no dependence on the results from varying \nfc\ aside from some mild trends at low flux density. This validates the robustness of the Fourier component model even when \nfc\ is larger than necessary. The more severely contaminated 8x-\planck\ mocks show similar results for completeness but a strong dependence of the false discovery rate on \nfc. As more Fourier components are fit to the data, fewer spurious sources are favored to absorb residual diffuse signal. 

\section{Measuring the extended Sunyaev-Zel'dovich effect in galaxy clusters}
\label{sec:szeffect}
To highlight the ability of \pcatde\ to disentangle components using spatial and spectral information, we apply our model to measure the thermal SZ effect toward massive galaxy clusters. The SZ effect describes the spectral distortion of CMB photons that are inverse Compton-scattered by electrons comprising hot gas in the intracluster medium (ICM). By measuring the SZ effect signal toward galaxy clusters and its dependence with frequency, one can probe the thermodynamics of the cluster ICM through the thermal SZ (tSZ) effect and its relativistic corrections (rSZ) \citep{SZeffect, Wright79, ItohrSZ, Chluba13} and cluster peculiar velocities through the kinematic SZ effect \citep[kSZ effect, e.g.,][]{Sheth2001,Sayers2016,Sayers2019}. At moderate redshifts, the SZ effect signal typically has an angular extent of several arcminutes, meaning it can be spatially correlated with CIB galaxies, foreground cirrus dust, cluster member galaxies, and potentially thermal dust emission associated with the cluster itself \citep{erler}. 

The SZ effect has a spectral energy distribution (SED) that rises with wavelength across the SPIRE bandpasses. However at 500 $\mu$m, that in which SZ has the largest surface brightness distortion, source blending is also more pronounced due to diffraction-limited optics. In contrast, at shorter wavelengths where the SZ effect signal is smaller in amplitude, CIB sources are more effectively detected and deblended. This is because the corresponding SPIRE maps have higher angular resolution and because the majority of observed CIB sources have blue spectra. A joint fit across all bands can incorporate these various properties in a consistent manner and can help to reliably extract the SZ component. Color priors can help enhance catalog inferences, with the caveat that unique sources in color space may become more difficult to recover \citep{Feder20}.

Identifying and separating these components can be challenging even when instrument noise is low, especially for single-dish sub-mm measurements for which individual point sources are difficult to separate from truly diffuse emission. While ancillary catalogs provide information about the potential positions of sub-mm detected sources (e.g., deep optical or mid-infrared catalogs), extrapolations of source SEDs over a large wavelength range are required to predict sub-millimeter flux densities. The number of external counterparts per SPIRE beam can be as high as thirty per SPIRE beam \citep{Roseboom10}, meaning some reduction of the external catalog is necessary if the SPIRE data are to be used to constrain the sub-mm flux densities. 

Measuring the SZ effect signal from SPIRE observations is a transdimensional task, because the field of faint, confused CIB sources is spatially correlated with the diffuse components. Within the formalism of probabilistic cataloging, samples are drawn from the marginalized posterior on surface brightness template amplitudes, \begin{equation}
P( \vec{A}_{SZ} |D) \propto \int P(\mathcal{C})P(\vec{A}_{SZ} | D, \mathcal{C})d\mathcal{C},
\end{equation}
where $\mathcal{C}$ denotes the full catalog space.


\subsection{Tests on mock galaxy clusters}

\pcatde\ is tested in this section on mock data based on a set of clusters previously observed by \emph{Chandra} and \emph{Herschel} observatories. \emph{Herschel}-SPIRE observed 56 galaxy clusters as part of the HerMES and HLS programs, with map depths of $\simeq 1$--2 \mjybm\ noise RMS \citep{Oliver2012,Egami2010}, which is subdominant to the SPIRE confusion noise which is $\sim$6 \mjybm\ at 250 $\mu$m \citep{Nguyen_2010}. Three clusters from this sample are chosen with properties listed in Table \ref{tab:clus_results}. For each cluster we compute an effective angular FWHM, $\theta_{\textrm{FWHM}}$, as the geometric mean of the cluster profile principal axes following image convolution of the cluster gas pressure profile with the SPIRE PSF. Our three clusters vary between 2.1 and 3.6 arcminutes and have temperatures spanning 8.3 and 17.3 keV, allowing us to probe a range of sizes and SZ effect amplitudes.

The procedure for generating mock cluster observations is detailed in \cite{Butler21}. In brief we use the same \texttt{B12} CIB model, combined with SPIRE noise realizations unique to the cluster observation with a mean noise RMS of $\sim$2 mJy beam$^{-1}$. The SZ signal component is modeled with a set of fixed templates $\textrm{I}^{\textrm{SZ}}_b$ with amplitudes \ASZb, convolved with the beam:
\begin{equation}
\lambda_{ij}^{b, SZ} = \mathcal{P}^b \circledast \left[ \textrm{A}_b^{\textrm{SZ}} \textrm{I}^{\textrm{SZ}}_b(x_i,y_j) \right].
\end{equation}
The same templates are then included in the forward model, i.e. $\lambda_{ij}^{b, SZ}$ is added to $\lambda_{ij}^b$ in Eq.~\ref{eq:gen_model}. The SZ effect signal is negligible at 250 $\mu$m, so we only fit SZ template amplitudes for SPIRE's 350 $\mu$m and 500 $\mu$m bands, denoted \ASZpmw\ and \ASZplw\ respectively. The morphology of the SZ profile is assumed to follow an elliptical generalized Navarro-Frenk-White (gNFW) profile \citep{Evans06}. This model is fit to ancillary Bolocam 140 GHz data, after which the best-fit profile is extrapolated and re-gridded to match SPIRE observations, following the method from \cite{Sayers2019}. We place no priors on the SZ template amplitudes nor on their colors, for the purpose of obtaining more data-driven constraints on the SZ effect signal. When one has a complete model for the signal considered (which is not the case for SZ spectral measurements), \pcatde\ is able to incorporate priors across diffuse components across bands. Cirrus realizations are not added to this set of mocks and no Fourier component model is included, though in reality a small fraction of clusters are observed through lines of sight with significant cirrus contamination exists. Fortunately, cirrus is well constrained by the high resolution 250$\mu$m data, for which the Fourier components can be fit simultaneously across bands assuming some color prior.

Galaxy clusters gravitationally lens background emission, which has the effect of deflecting and magnifying light from CIB sources. While surface brightness is conserved by lensing, the net effect after removing bright detected sources is a surface brightness deficit near the center of the cluster. This was first measured in \cite{Zemcov_lens} in four clusters, and can bias measurement of the SZ signal because the two can be highly spatially degenerate. Bias due to lensing is estimated and corrected in \cite{Sayers2019} and \cite{Butler21}, however for simplicity the results shown in this work use unlensed mock CIB realizations. In the absence of lensing in the observed data, our forward model is fully specified. 

\subsection{SZ results}

We test \pcatde\ on mock observations toward galaxy cluster RXJ 1347.5-1145, which has been the subject of numerous SZ studies \citep{Pointecouteau1999, Komatsu2001, Kitayama2004, Zemcov2012, Sayers2016_RXJ, Kitayama2016}, including one that uses \pcatde\ \citep{Butler21}. To derive constraints on cluster properties like the temperature of gas comprising the intracluster medium (ICM) or cluster peculiar velocity, multi-wavelength data from several instruments (e.g., \textit{Bolocam}, \textit{Planck}, \textit{Chandra}, {\it Hubble Space Telescope} ({\it HST})) are commonly employed, however in this work we focus on surface brightness measurements from SPIRE data alone.

\subsubsection{Convergence of SZ parameters}
To validate that the SZ template amplitude parameters are converged, we compute the Gelman-Rubin statistic, or the potential scale reduction factor (PSRF, also known as $\hat{R}$), from several Markov chains run on the same data. Twenty MCMC walkers are independently initialized and run on a single mock cluster realization of RXJ 1347.5-1145 for 4000 thinned samples, with the second half of each chain used to compute $\hat{R}$. We estimate $\hat{R}=1.07$ and $\hat{R}=1.08$ for \ASZpmwhat\ and \ASZplwhat, respectively, suggesting the chains are well mixed.

\subsubsection{Component separation}
\begin{figure*}
    \centering
    \includegraphics[width=\linewidth]{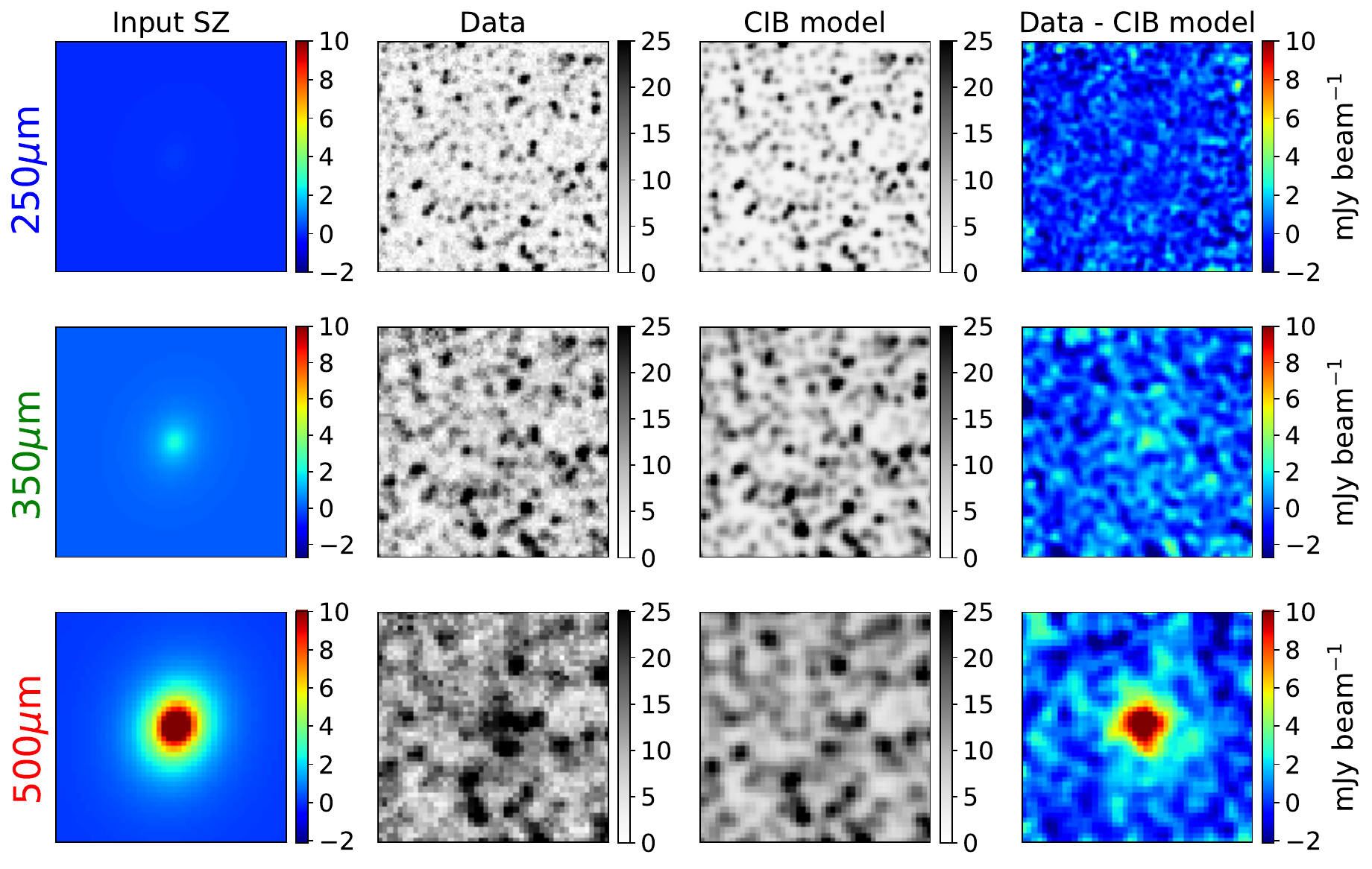}
    \caption{$10\arcmin \times 10\arcmin$ cirrus-free mock SPIRE observations toward galaxy cluster RX J1347.5-1145. The columns show (going from left to right) the input SZ effect signal, observed maps, median CIB model, and difference between observed maps and CIB model. By fitting both SZ and CIB simultaneously we can account for the presence of sub-mm point sources without overfitting the underlying SZ signal. The maps in the rightmost column are shown for visualization purposes only.}
    \label{fig:sz_resid}
\end{figure*}

An advantage of \pcatde\ for this application is that, by modeling all components simultaneously one mitigates parts of the SZ signal being apportioned to point sources and vice versa. This can be understood upon visual inspection in Fig. \ref{fig:sz_resid}, where the observed cluster field, the best fit CIB model, and the residual between the two are plotted. Even when the input SZ signal has a small signal to noise ratio (for example, at 350$\mu$m), or is heavily confused as seen at 500$\mu$m, \pcatde\ is able to reliably separate the underlying signal from contaminants. Unmodeled point source emission can be seen as well in the residual maps, meaning confusion noise remains a significant systematic in the surface brightness measurement.

\subsubsection{Sensitivity to cluster properties}
\begin{figure*}[t]
    \centering
    \includegraphics[width=\linewidth]{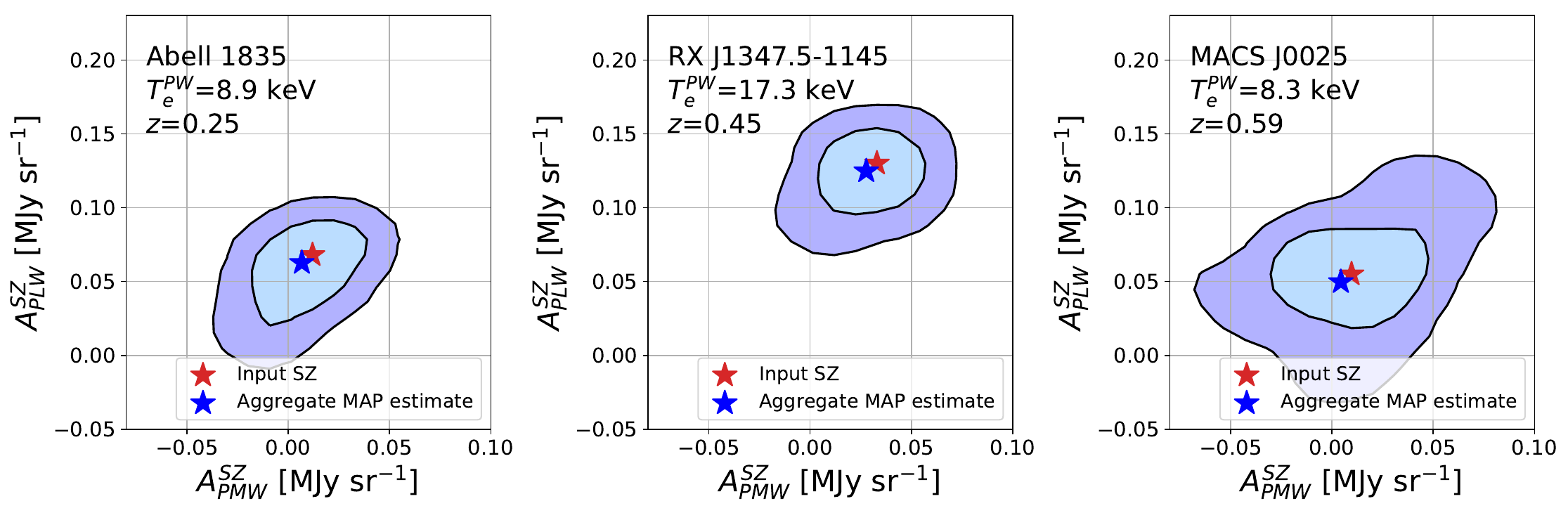}
    \caption{Recovered posteriors on \ASZb\ surface brightness parameters for clusters of varying redshift/angular extent. The cluster realizations corresponding to these results are cirrus-free, and input SZ surface values for each cluster (marked by red stars) are determined from a combined analysis of \emph{Chandra} and Bolocam data. 1$\sigma$ and 2$\sigma$ density contours are plotted for each cluster.}
    \label{fig:diff_asz}
\end{figure*}

For each of the three clusters listed in Table \ref{tab:clus_results}, the same set of twenty CIB + instrument noise realizations are used to generate mock cluster observations, after which samples from each set of twenty chains run on the data are aggregated and plotted in Figure \ref{fig:diff_asz}. Computing this full distribution allows us to quantify systematic uncertainty associated with the CIB, identify any consistent biases and assess the sensitivity of our results to details of the cluster itself.

While source confusion can have large effects on the recovered SZ effect signal, the recovered surface brightness estimates are fairly unbiased over several CIB realizations. The estimated maximum \emph{a posteriori} (MAP) values and 68 per cent credible intervals for \ASZpmw\ and \ASZplw\ are compared with input surface brightnesses in Table \ref{tab:clus_results}. The mean bias is $\lesssim-0.3\sigma$ for both \ASZpmw\ and \ASZplw, and this bias is consistent across our three clusters. This implies it is primarily correlated with the common CIB mocks used to make each set of cluster observations. In general the derived uncertainties do not vary significantly from cluster to cluster, however a more thorough investigation of uncertainties from a larger sample of clusters may reveal trends with respect to gas temperature, angular extent, etc. We find that MACS J0025, the cluster with lowest gas temperature (i.e., smallest SZ distortion) and smallest angular extent, has larger uncertainties by $\sim$40 and 20 per cent for PMW and PLW, respectively.

\begin{table*}[]
    \centering
    \begin{tabular}{c|c|c|c|c|c|c|c}
        Cluster name & Redshift & T$_{\textrm{PW}}$ & $\theta_{\textrm{FWHM}}$ & Input \ASZpmw\   & \ASZpmwhat\  & Input \ASZplw & \ASZplwhat\ \\
        & & [keV] & [arcmin] & [MJy sr$^{-1}$] & (MAP, 68\% C.I.) & & \\
        \hline
        Abell 1835 & 0.25 & 8.9 & 3.6 & 0.012 & $0.007^{+0.016}_{-0.014}\quad(0.5\sigma)$ & 0.068 & $0.063^{+0.014}_{-0.026}\quad(2.4\sigma)$\\
        RX J1347.5-1145 & 0.45 & 17.3 & 2.5 & 0.033 & $0.028^{+0.016}_{-0.013}\quad(2.2\sigma)$& 0.130 & $0.125^{+0.014}_{-0.020}\quad(6.3\sigma)$\\
        MACS J0025 & 0.59 & 8.3 & 2.1 & 0.010 & $0.004^{+0.033}_{-0.018}\quad(0.2\sigma)$&  0.055 & $0.050^{+0.031}_{-0.018}\quad(2.2\sigma)$
    \end{tabular}
    \caption{Input and recovered SZ effect surface brightnesses for the three test clusters Abell 1835, RXJ 1347 and MACS J0025. These clusters have been observed by \emph{Chandra}, Bolocam and SPIRE, with pressure-weighted temperatures derived from \emph{Chandra} observations. Surface brightnesses are in units of MJy sr$^{-1}$ and uncertainties are reported using the highest posterior density intervals. The significance of each SZ detection is also computed assuming Gaussian uncertainties.}
    \label{tab:clus_results}
\end{table*}

\section{Summary and Outlook}

In this work we have considered a variety of inference tasks that rely on effective separation of point-like and diffuse signals. This is done by extending the forward modeling framework of probabilistic cataloging (PCAT) with a flexible, template-based model for diffuse signals, which results in efficient separation of CIB emission from diffuse Galactic cirrus. Our point source completeness limits (\S \ref{sec:ptsrc}) are robust to additional diffuse emission, degrading by less than $\sim 20\%$ at nearly all flux densities and both noise levels. The purity degrades by a larger amount in the presence of prominent diffuse signals, and we find this degradation correlates with the local signal curvature. At the population level, \pcatde\ enables robust recovery of number counts across a range of cirrus foreground levels. For our low noise ($\sigma_{inst} = 1$ \mjybm) case we obtain consistent flux density number counts across all levels of injected cirrus for $S_{250}>25$ mJy, and the counts in cases up to 4x-\planck\ are consistent for $S_{250}>12$ mJy.

In our second application of \pcatde\ we demonstrate that the faint, spatially extended SZ effect signal can be recovered in the presence of CIB galaxy contamination along the same line of sight. Averaged over several mock CIB realizations, the mean recovered SZ surface brightnesses are unbiased at the $\lesssim 0.3\sigma$ level, with uncertainties dominated by confusion noise. \pcatde\ was recently used to measure the gas temperature of galaxy cluster RX J1347.5-1145, to which the SZ spectrum is sensitive through relativistic effects \citep{Butler21}. That work found a temperature of \Ttwofive$=22.4^{+10.6}_{-12.0}$ keV, consistent with the X-ray measured \Tpwxtwofive\ = 17.3 keV. The results from this work further demonstrate that similar measurements of the SZ effect should be robust for a collection of different cluster profiles and gas temperatures. When a spatial template for the diffuse signal is available it can be easily incorporated into the forward model as shown in \S 6, however more detailed signal parameterizations may be used when appropriate.

Surveys with strict requirements on  photometric accuracy may place stringent cuts on sources based on estimated signal to noise ratio, or excise regions with pronounced diffuse contamination. In addition to a loss of information, the situation can be especially problematic when uncertainties due to diffuse signals are underestimated. We anticipate the tools from this work can both expand the sample size of ``usable" sources in astronomical catalogs when source confusion and diffuse foregrounds are prominent. 

While the Fourier component model performs well for the examples considered in this work, there are limitations on the types of signals it can reconstruct effectively. In particular, structures comparable to or smaller than the PSF FWHM may be more difficult to model with Fourier components, as suggested by results on cirrus-dominated maps (see Fig. \ref{fig:rms_vs_nfc}). More flexible generative models may be able to capture nonlinear structures such as filaments, as have been demonstrated on CMB data to model foregrounds \citep{gendust, gendust2}. Under the assumption that the color of the diffuse signal component is constant over a given field of view, it is effective to model the Fourier components in several bands with a simple linear scaling factors, incorporating color prior information when appropriate. Position-dependent color variations may be non-negligible for some observations, which can be addressed either by processing smaller regions with fixed color or by incorporating a model for color variations. The formalism of Bayesian hierarchical modeling permits for more detailed extensions of the forward model, in a way that reflects an appropriate level of knowledge. By the same token one should always be careful in characterizing the effect of priors, both explicitly specified and those implicit to the method \citep[][]{redshiftpz}, on a given inference task.

\pcatde\ is publicly available on Github (\url{github.com/RichardFeder/pcat-de}), with corresponding documentation\footnote{\url{pcat-de.readthedocs.io}} and examples demonstrating applications from this work. 

    
    

\facilities{\emph{Herschel}, CS0, CX0, \emph{Planck}} 
\software{\texttt{corner}, \texttt{matplotlib}, \texttt{numpy}, \texttt{Lion}}

\appendix

\section{Fourier component marginalization}
\label{sec:fc_marg}
To accelerate the burn-in phase of sampling, we apply a series of linear marginalization steps for the set of Fourier component templates, the number of which are fixed \emph{a priori}. Let $\mathcal{F}_{ij}$ be the $i^{th}$ pixel of the $j^{th}$ 
Fourier component template, and let $\beta_j$ be the $j^{th}$ Fourier component's amplitude. Then $\pmb{\mathcal{F}\beta}$ is a column vector with the flux in each pixel from each Fourier component. Let $\pmb{\Sigma}$ be a diagonal matrix where $\Sigma_{ii}$ is the variance in pixel $i$ and $K$ the corresponding data vector. Then the $\pmb{\beta}$ that minimizes the chi-squared statistic $\chi^2(\pmb{\beta})$ is given by the Moore-Penrose inverse:

\begin{align}
\label{marg_noreg}
    \hat{\pmb{\beta}} &= (\pmb{\mathcal{F}}^{T}\pmb{\Sigma}^{-1}\pmb{\mathcal{F}})^{-1}\pmb{\mathcal{F}}^{T}\pmb{\Sigma}^{-1}K.
\end{align}

While the linear inversion from equation \eqref{marg_noreg} minimizes $\chi^2$ with respect to the data, in practice the Fourier component parameters are driven to local minima that are difficult to leave in the sampling phase. In addition, because the marginalization procedure is only applied to Fourier component templates, the MAP estimates obtained at the beginning of burn-in are conditioned on an unconverged point source model. When used at the beginning of the sampler to obtain an initial guess of the diffuse model, the MAP estimate is biased due to unmodeled point sources. To prevent divergence of the Moore-Penrose inverse, the solution is regularized by imposing a prior on the Fourier component coefficients. This is done through ridge regression, which penalizes the loss function with the $\ell_2$ norm of the component amplitudes:
\begin{equation}
 \hat{\pmb{\beta}}_{ridge} = (\pmb{\mathcal{F}}^{T}\pmb{\Sigma}^{-1}\pmb{\mathcal{F}} + \pmb{\sigma} \pmb{I})^{-1}\pmb{\mathcal{F}}^{T}\pmb{\Sigma}^{-1}K.
 \label{eq:ridge}
\end{equation}
The vector $\pmb{\sigma}$ acts as a Gaussian prior on the fluctuation amplitude of each component. In \texttt{PCAT-DE}, $\pmb{\sigma}$ is inversely proportional to the power spectrum of the underlying signal evaluated at the scale of each Fourier component. An iterative scheme is implemented in \texttt{PCAT-DE} in which the Fourier components are repeatedly fit to the residual of the data and point source model. The terms in equation \eqref{eq:ridge} only need to be computed once up front and then stored for fast evaluation with the residual data vector $K$. For a $100\times100$ pixel image fit using a \nth{15} order Fourier component model, computing $\pmb{\hat{\beta}}$ takes 8 ms per evaluation (with pre-computed quantities), accelerating the burn-in MCMC phase of the fitting routine. Extending the marginalization procedure to include source fluxes should be possible but is left to future work.


\section{Parsimony prior in the weakly non-asymptotic limit}
\label{sec:parsimony_prior_app}
In probabilistic cataloging, goodness of fit is enforced with a delta log-prior proportional to the number of point source parameters being added or removed from the model. This reflects the fact that adding parameters to a model will, on average, improve the log-likelihood of a reconstructed signal. When the number of parameters is much smaller than the dimension of the data, the expected improvement in the log-likelihood is 1/2 per additional degree of freedom. However, this asymptotic result may be ill-suited in the limit of severely confused observations. 

We consider the F-statistic \citep{fdist}, which models the significance of a model's improved fit to data using Snedecor's F-distribution. Let $\mathcal{M}_1$ be an initial model with $p_1$ parameters and $\mathcal{M}_2 = \mathcal{M}_1\cup \delta$, the union of the initial model and $\delta$ with $p_2 = p_1 + p_{\delta}$ parameters. The F-statistic is expressed in terms of the chi-squared statistic, the number of parameters associated with each model and the length of the data vector:
\begin{equation}
    F = \frac{\chi_1^2-\chi_2^2}{p_{\delta}}\frac{N_{pix}-p_2}{\chi_2^2}
\end{equation}
Then one can rearrange terms to relate the delta chi-squared to the F-distribution with $p_{\delta}$ and $N_{pix}-p_2$ degrees of freedom:
\begin{equation}
    \Delta\chi^2 \sim \frac{\chi_2^2 p_{\delta}}{N_{pix}-p_2}F(p_{\delta}, N_{pix}-p_2)
\end{equation}
The expectation of this F-distribution is $\frac{N_{pix}-p_2}{N_{pix}-p_2-2}$, and so the expected delta log-likelihood is
\begin{equation}
    \langle \Delta\log \mathcal{L} \rangle = \frac{1}{2}\frac{\chi_2^2 p_{\delta}}{N_{pix}-p_2-2}.
    \label{eq:dlogL}
\end{equation}

\begin{figure}[h]
    \centering
    \includegraphics[width=0.7\linewidth]{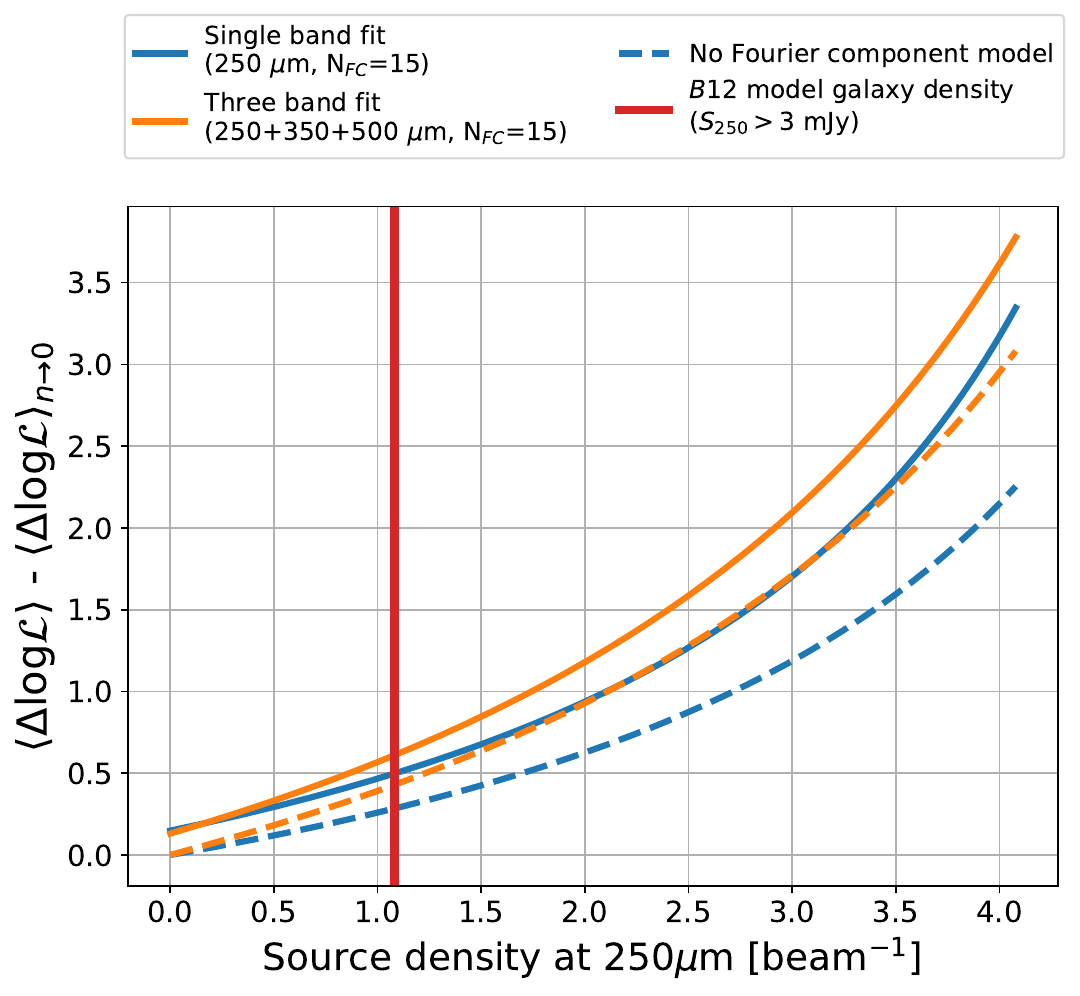}
    \caption{Expected improvement in log-likelihood from adding a point source to the model relative to sparse limit, $\langle \Delta \log \mathcal{L}\rangle_{n\to0}$, as a function of source density. Three parameters are added to the model per source in the single band case, while each additional band included in a joint fit increases the number of parameters per source by one. Solid and dashed curves are for results with and without Fourier components, respectively.}
    \label{fig:parsimony_prior}
\end{figure}

In Figure \ref{fig:parsimony_prior} we plot Eq. \ref{eq:dlogL} for both single- and three-band cases assuming good fits to the data, (i.e., $\chi_2^2 \approx N_{pix}$). This is done as a function of source density, ranging from zero (the ``sparse limit") to four sources per beam. For both curves, the larger improvement relative to the asymptotic limit is expected -- correlations between model parameters become non-negligible, meaning the model components can conspire to produce a better reconstruction of the data. These results further suggest that a three-band fit to \emph{Herschel}-SPIRE data becomes more susceptible to overfitting as the source density increases. This is explained by the poorer angular resolution of the 350 and 500 $\mu$m SPIRE maps -- while there are additional pixels to constrain the model, these are outnumbered by the additional parameters required to model source fluxes in these bands, leading to more overfitting. This would not be the case if the resolution were the same across maps, for example in a joint fit of several 250 $\mu$m observations of the same field. A simpler point source parameterization, for example colors modeled by a single temperature, would reduce the number of additional parameters and thus the amount of overfitting, but care would be needed to manage the transition between sources whose observed spectra trace the black-body peak (i.e., sources with temperature-driven colors) and sources with more complicated SEDs. 

We use the expected number density of SPIRE sources with $S_{250} > 3$ mJy based on the \cite{Bethermin2012} model ($\sim 1$ beam$^{-1}$) to predict the expected delta log-likelihood relative to the sparse limit, which informs our choice of scaling parameter $\rho$ defined in \eqref{eq:parsimony_prior}. Additional parameters for the Fourier component model modify the parsimony prior as well, as can be seen comparing the solid and dashed curves in Fig. \ref{fig:parsimony_prior}.

\section{Source/background Covariance}
\label{sec:cov_ptsrc_bkg}
Quantifying the covariance between point sources and a generic foreground/background helps to inform the proposal kernel widths for model components within \texttt{PCAT-DE}. As detailed in \cite{psf2020}, the uncertainty of a uniform background $b$ in the presence of a single point source with flux density $f$ can be calculated by computing the 2$\times$2 Fisher information matrix and its inverse:
\begin{equation}
    \sigma_b^2(\theta_{ML}) = \frac{-\partial_f^2 \ln \mathcal{L}(\theta_{ML})}{\partial_b^2\ln\mathcal{L}(\theta_{ML})\partial_f^2\ln\mathcal{L}(\theta_{ML})-(\partial_b\partial_f\ln\mathcal{L}(\theta_{ML}))^2} = \frac{\sigma^2}{A - A_{psf}}
\end{equation}
where $A = nm$ and $A_{psf} = \left(\sum_i^{N_{pix}}p_i(x,y)^2\right)^{-1}$ is the effective PSF area in pixels. This means that a larger effective PSF increases the corresponding uncertainty on $b$. 

We can generalize the previous result to include $n$ sources in the covariance matrix. Let us assume that our source fluxes are independent of one another, i.e. mixed-derivative terms $\partial_{f_i}\partial_{f_j}\ln \mathcal{L}^* = 0$, where the maximum log-likelihood is abbreviated $\ln\mathcal{L}^* = \ln\mathcal{L}(\theta_{ML})$. Our Fisher matrix can then be written as the following:
\begin{equation*}
    \mathcal{F}_{(\lbrace f_i\rbrace_{i=1}^n, b)}(x,y) = \kbordermatrix{ & f_1 & f_2 & \hdots & f_n & b \\ f_1 & \partial_{f_1}^2\ln\mathcal{L}^* &  &  & & \partial_{f_1}\partial_b \ln\mathcal{L}^*\\ f_2 &  & \partial_{f_2}^2 \ln\mathcal{L}^*&  &  & \partial_{f_2}\partial_b \ln\mathcal{L}^* \\ \vdots &  &  & \ddots & & \vdots \\ f_n &  &  &  & \partial_{f_n}^2 \ln\mathcal{L}^* & \partial_{f_n}\partial_b \ln\mathcal{L}^* \\ b & \partial_b \partial_{f_1}\ln\mathcal{L}^* & \partial_b \partial_{f_2}\ln\mathcal{L}^* & \hdots & \partial_b\partial_{f_n}\ln\mathcal{L}^* & \partial_b^2\ln\mathcal{L}^*}
\end{equation*}
Assuming the diagonal elements in the above arrowhead matrix are non-zero, the inverse is a rank-one modification of a diagonal matrix:
\begin{equation}
    C_{(\lbrace f_i\rbrace_{i=1}^n, b)}(\theta_{ML}) = [\mathcal{F}_{(\lbrace f_i\rbrace_{i=1}^n, b)}(\theta_{ML})]^{-1} = \left[\begin{array}{cc}D^{-1} & \\ & 0\end{array}\right] + \rho u u^T
    \label{eq:cov_srcs}
\end{equation}
where
\begin{align*}
    D &= \text{diag}(\partial_{f_1}^2\ln \mathcal{L}^*,...,\partial_{f_n}^2\ln\mathcal{L}^*); \\
    u &= \left[\begin{array}{cc} D^{-1}z & -1\end{array}\right]^{T}; \\
    z &= \left[\begin{array}{cccc}\partial_{f_1}\partial_b \ln\mathcal{L}^* & \partial_{f_2}\partial_b \ln\mathcal{L}^* & \hdots & \partial_{f_n}\partial_b \ln\mathcal{L}^*\end{array}\right]^{T}; \\
    \rho &= \frac{1}{\alpha - z^T D^{-1}z}; \\
    \alpha &= \partial_b^2 \ln \mathcal{L}^*.
\end{align*}
Computing the covariance matrix from equation \eqref{eq:cov_srcs}, and evaluating the diagonal background element, one obtains
\begin{equation}
    \sigma_b^2(\theta_{ML}) = \frac{\sigma^2}{A - nA_{PSF}}.
\end{equation}
This follows the intuition that as the number of sources in a given image increases, the effective number of pixels that contribute to constraining the background normalization decreases. In more realistic situations where the correlations between sources are taken into account, the uncertainties on $b$ will in general be larger. While we consider uncertainties on the mean normalization, a discussion of uncertainties on a generic sky model is presented in Appendix C of \cite{psf2020}.

\bibliography{references}{}
\bibliographystyle{aasjournal}

\end{document}